\documentclass[%
 reprint,
 amsmath,amssymb,
 aps,prl
]{revtex4-1}

\usepackage{graphicx}
\usepackage{dcolumn}
\usepackage{bm}

\usepackage{color}

\begin{document}

\preprint{APS/123-QED}

\title{Intrinsic emergence of Majorana modes in Luttinger $j=3/2$ systems}

\author{Julian Benedikt Mayer}
\author{Miguel A. Sierra}%
\author{Ewelina M. Hankiewicz}
 \affiliation{Institute for Theoretical Physics and Astrophysics and W\"urzburg-Dresden Cluster of Excellence ct.qmat, University of W\"urzburg, Am Hubland, 97074 W\"urzburg, Germany}
 

\date{\today}

\begin{abstract}
We analyze theoretically two different setups for s-wave superconductivity (SC) proximitized $j=3/2$ particles in Luttinger materials that are able to host Majorana bound states (MBSs). First, we consider a one-dimensional SC wire with intrinsic bulk inversion asymmetry (BIA). In contrast to wires, modeled by a quadratic dispersion with Rashba spin-orbit coupling, there are two topological phase transitions in our systems at finite magnetic fields. Second, we analyze a two-dimensional Josephson junction on the Luttinger model finding a topological region even in the absence of BIA and Rashba spin-orbit couplings. 
This originates from the hybridization of the light and heavy hole bands of the $j=3/2$ states in combination with the SC pairing. As a consequence, both systems can be driven into a topological phase hosting MBSs. Hence, we predict that MBSs form in any SC proximitized Josephson junction on 2D Luttinger materials by the application of magnetic field alone. This opens a new avenue for the search of topological SC.
\end{abstract}

\maketitle


\section{\label{sec:Introduction}Introduction}

Remarkable interest in Majorana bound states (MBSs) has arisen in the last decades \cite{Alicea2012,Beenakker2013,Lutchyn2018,Prada2020,Aguado2020}. In condensed-matter systems, they manifest as zero-energy modes appearing at the boundaries of a topological superconductor (SC) \cite{Kitaev2003,Leijnse2012}. Since they come in pairs to form a fermionic state, they acquire nonlocal properties which topologically protect them from decoherence \cite{Kitaev2003,Fu2008,Goldstein2011,Leijnse2012,Aguado2020,Prada2020}. For this reason, MBSs are ideal candidates for topological quantum computation. Consequently, several experiments attempt to detect signatures of MBSs \cite{Mourik2012,Deng2018,Lutchyn2018,Fornieri2019,Ren2019}.

It has been predicted that a semiconducting nanowire with s-wave proximitized SC and spin-orbit coupling (SOC) can host MBSs at its boundaries \cite{Lutchyn2010,Oreg2010,Wakatsuki2014,Szumniak2017,Prada2020,Kharitonov2020}. 
These appear in a topological phase after a Zeeman field, perpendicular to the SOC field, inverts the SC gap.
Similarly, a two-dimensional Josephson junction with perpendicular Zeeman and SOC fields is also able to host MBSs at the boundary between the normal region and the vacuum \cite{Peng2016,Pientka2017,Fornieri2019,Ren2019,Scharf2019}.
These MBSs appear, if the double degenerate Andreev bound states (ABS) split at finite Zeeman fields, giving rise to a topological regime between two crossings. The topological gap in the ABS spectrum protects the MBSs against perturbations.

\begin{figure}
    \centering
    \includegraphics[width=\linewidth]{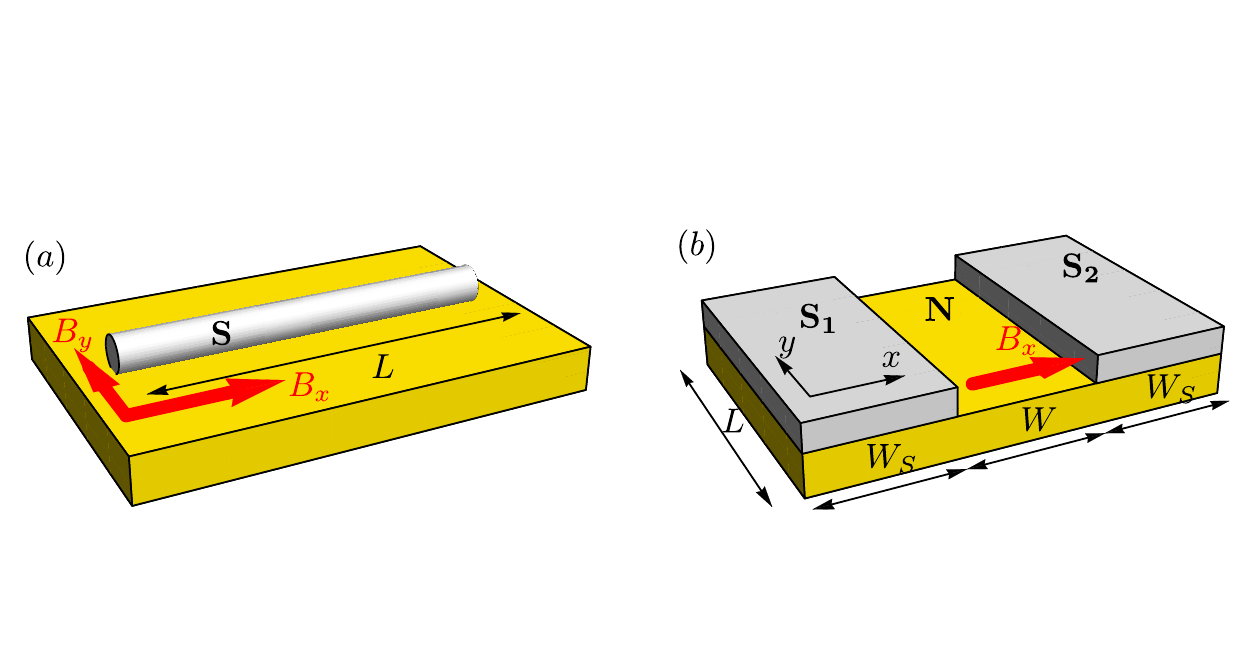}
    \caption{Sketch of the setups: (a) A one-dimensional Luttinger wire of length $L$ in contact with an s-wave superconductor and in the presence of a Zeeman field $B_x$ or $B_y$. (b) A two-dimensional Josephson junction of length $L$ with a Zeeman field applied in a normal region of width $W$. Two s-wave superconductors of width $W_S$ and phases $\phi_{S_1}=-\phi/2$ and $\phi_{S_2}=\phi/2$, respectively, separated by a normal regime $N$ on top of the Luttinger material.}
    \label{fig:setup}
\end{figure}

In this Letter, we analyze these setups for the $j=3/2$ states of bulk materials like HgTe, $\alpha$-Sn, or half-Heusler compounds, which exhibit a quadratic band touching at the $\Gamma$-point around the Fermi energy. These so-called quadratic nodal semimetals can be described within the 4-band Luttinger model (LM) \cite{Ruan2016,Kharitonov2017}. 
Luttinger materials display a rich variety of topological phases induced by perturbations, i.~e., in the presence of strain they are 3D topological insulators \cite{Brune2011,Brune2014,Baum2014} or topological semimetals \cite{Morimoto2014,Kondo2015,Ruan2016,Zhang2018,Ghorashi2018}, with local attractive electron-electron interaction they transform into superconducting phase \cite{Boettcher2016,Boettcher2018}, and they can show higher spin SC pairings \cite{Brydon2016,Kim2018,Roy2019,Kim2020,Dutta2021,Timm2021,Bahari2022}. 
Our models are appealing for two reasons: on the one hand, they give a more realistic description of materials. On the other hand, the LM can be derived from the $k\cdot p$ model in which spin-orbit interactions, especially bulk inversion asymmetry (BIA) terms, are already taken into account \cite{Winkler2003}. Hence, MBSs could emerge intrinsically with no extra implementations of spin-orbit interactions in the setup. 

Specifically, we show that in a 1D LM wire, MBSs emerge from the interplay between intrinsic BIA and the Zeeman field. 
Further, in contrast to the semiconducting Rashba nanowires, the topological phase in our systems is limited to a finite Zeeman field range.
Moreover, we predict that for 2D Josephson junctions on Luttinger materials, MBSs still appear due to the intrinsic SOC originating from the mixing between light hole (LH) and heavy hole (HH) bands.
Therefore, in Josephson junctions on 2D materials described by $j=3/2$ particles no extra SOC is necessary to generate MBSs.

\section{\label{sec:Model}Theoretical models}

We use the 4-band LM \cite{Kohn1954} to describe the $j=3/2$ electronic states of our systems:
\begin{eqnarray} \label{Eq:LuttHam}
    \hat{H}_L(\bm{k}) &=& \alpha_0 \bm{k}^2\hat{1}_4 + \alpha_z \hat{M}_z(\bm{k}) + \alpha_\square \hat{M}_\square(\bm{k}) - \mu \hat{1}_4,
\end{eqnarray}
with $\hat{M}_z(\bm{k}) = \tfrac{5}{2}\bm{k}^2 \hat{1}_4 - 2(\bm{k}\cdot \hat{\bm{J}})^2$ and $\hat{M}_\square(\bm{k})= k_x^2\hat{J}_x^2 + k_y^2 \hat{J}_y^2 -\tfrac{2}{5}(\bm{k}\cdot \hat{\bm{J}})^2 - \tfrac{1}{5}\bm{k}^2 \hat{\bm{J}}^2$. Here, $\bm{k} = (k_x,k_y,0)$ is the momentum, $\bm{\hat{J}}=(\hat{J}_x,\hat{J}_y,\hat{J}_z)$ are the $j=3/2$ spin matrices, and $\alpha_0$, $\alpha_z$, and $\alpha_\square$ are material-specific parameters related to the effective masses of the bands while $\mu$ is the chemical potential.
We emphasize that the $\alpha_z$ and $\alpha_\square$ terms involve an intrinsic symmetric SOC in the system for at least two dimensions \cite{Brydon2016}. 

Using the Nambu basis $(c_{\frac{3}{2}}, c_{\frac{1}{2}}, c_{-\frac{1}{2}}, c_{-\frac{3}{2}}, -c_{-\frac{3}{2}}^\dagger, c_{-\frac{1}{2}}^\dagger, -c_{\frac{1}{2}}^\dagger, c_{\frac{3}{2}}^\dagger)$, where $c_{j_z}^\dagger$ ($c_{j_z}$) are the creation (annihilation) operators of particles with z-componenet of total angular momentum $j_z$, we get the Bogoliubov-de Gennes (BdG) Hamiltonian describing a 1D SC wire [Fig.~\ref{fig:setup}(a)]
\begin{equation}
    \hat{H}_W(k_x)= \left. \hat{\tau}_z \hat{H}_L(k_x)\right|_{k_y=0} + \hat{\tau}_z \hat{H}_{BIA}(k_x) + \Delta \hat{\tau}_x \hat{1}_4 + B_y \hat{\tau}_0 \hat{J}_y,
\end{equation}
where $\bm{\hat{\tau}}=(\hat{\tau}_x,\hat{\tau}_y,\hat{\tau}_z)$ are the Pauli matrices in Nambu space. The second term corresponds to the BIA $\hat{H}_{\mathrm{BIA}}(k_x) = \beta k_x \{\hat{J}_x, \hat{J}_y^2-\hat{J}_z^2 \}$, where $\beta$ is the BIA strength and $\{\ldots\}$ is the anticommutator. We identify $\hat{H}_{\mathrm{BIA}}$ as an intrinsic source of SOC interactions in any semimetal of the $\mathbf{T}_d$ tetrahedral symmetry group. The proximitized SC is represented by the induced s-wave pairing potential $\Delta$. In addition, a Zeeman term $B_y$ is applied perpendicularly to the BIA field.

For the 2D Josephson junction [Fig.~\ref{fig:setup}(b)], we consider the Hamiltonian 
\begin{equation}
 \hat{H}_{JJ}(\bm{k})=\hat{\tau}_z \hat{H}_L(\bm{k}) +\hat{H}_\Delta^{JJ}+ \hat{\tau}_0\hat{H}_Z^{JJ}.
\end{equation}
In this setup, two SCs ($S_1$ and $S_2$) are separated by a non-SC (N) region of width $W$. Therefore, the SC coupling takes the form $\hat{H}_\Delta = \Delta \Theta (|x|-W/2) \left[ e^{i\phi(x)} \hat{\tau}_++ e^{-i\phi(x)} \hat{\tau}_-\right]\hat{1}_4$ where $\Theta(x)$ is the Heaviside function, $\phi(x)=(\phi/2) \mathrm{sgn}(x)$ is the SC phase, and $\hat{\tau}_\pm = (\hat{\tau}_x\pm i\hat{\tau}_y)/2$. Additionally, the Zeeman field is only applied in the normal region such that $\hat{H}_Z^{JJ} = \Theta(W/2 - |x|)B_x \hat{J}_x$.

In the presence of a Zeeman field, our systems have broken time-reversal symmetry and conserved particle-hole symmetry. Therefore, they are in the symmetry class D \cite{Schnyder2008,Ryu2010} which is categorized by the topological invariant $\mathcal{Q}=\det(r)$ in 1D, where $r$ is the reflection matrix \cite{Akhmerov2011, Fulga2011}. 
We notice that, even though the Josephson junction is a 2D system, the normal region in which the MBSs appear can be considered as a quasi-1D wire since the MBS are localized along the $y$-direction only.
Therefore, we obtain the topological invariant of both setups by employing the method for 1D explained in Ref.~\cite{Fulga2011}.

The solutions to the scattering problem and BdG equation were calculated numerically using the Kwant package \cite{Kwant}. 

\begin{figure}[t!]
    \centering
    \includegraphics[width=\linewidth]{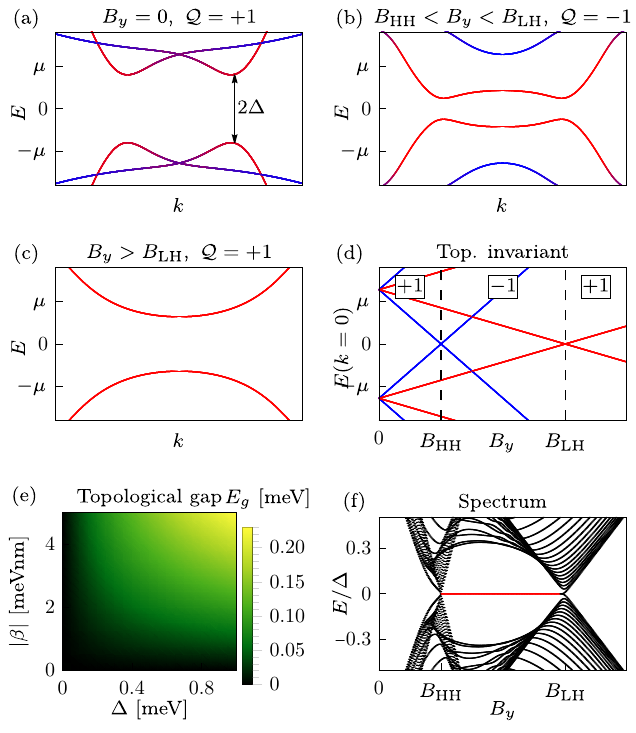}
    \caption{(a)-(d) Bulk bandstructure calculations of an infinite wire with proximitized superconductivity and Zeeman field $B_y$ [see Fig.~\ref{fig:setup}(a)], using the parameters of HgTe $(\alpha_0,\alpha_z,\alpha_\square)=(16.58,9.36,-1.6)\frac{\hbar^2}{2m}$ and $\beta=-4.31\;\mathrm{meV nm}$ \cite{Novik2005, Cardona1986}. We set the chemical potential at $\mu = 0.25 \; \mathrm{meV}$ and $\Delta = 0.2 \; \mathrm{meV}$. The color code indicates the character of the bands being light-hole (red) or heavy-hole (blue) like. The insets in (d) specify the topological invariant $\mathcal{Q}$. The critical magnetic fields $B_{\mathrm{HH}}$ and $B_{\mathrm{LH}}$ are given in Eq.~(\ref{Eq:Bcritical}). (e) Size of the minimum gap for all momenta as a function of the superconducting gap $\Delta$ and the strength of the bulk inversion asymmetry $\beta$ in the topological phase at $B_y=(B_{HH}+B_{LH})/2$. (f) Bandstructure of a wire with finite length $L$ as a function of $B_y$ with a system size much larger than the localization length of the MBS ($L\gg \lambda$). States localized at the boundaries of the wire are indicated with red color, showing the Majorana bound states in the topologically nontrivial region.}
    \label{fig:1}
\end{figure}

\section{\label{sec:SCwire} Superconducting wire}

\begin{figure*}[t!]
    \centering
    \includegraphics[width=\linewidth]{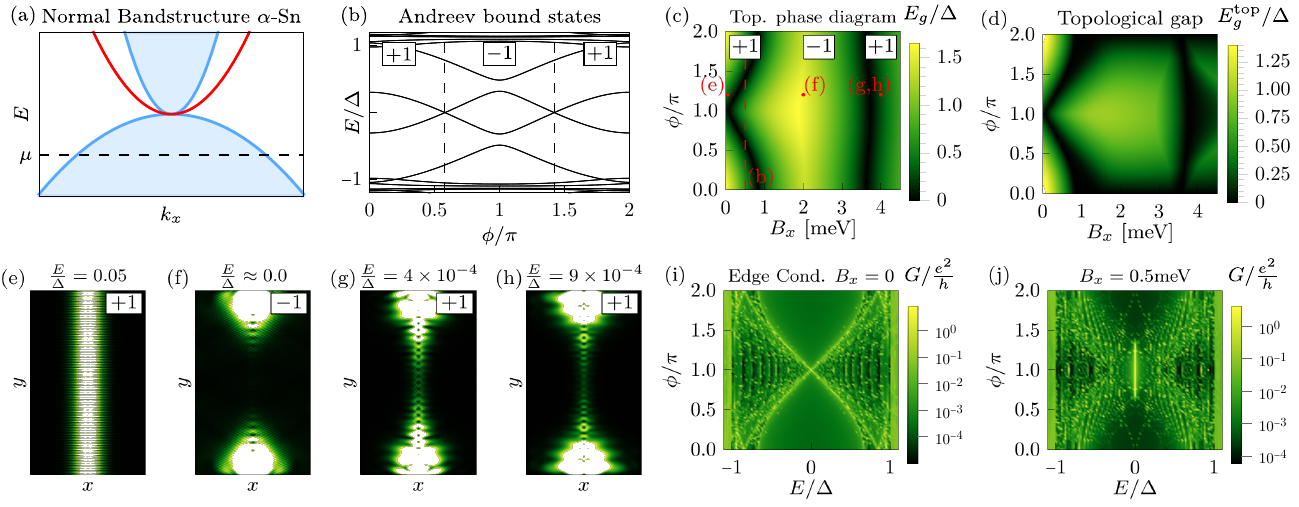}
    \caption{(a) Bandstructure of the non-superconducting semi-infinite 2D semimetal regime with the bulk continuum (blue shaded) and the edge states (red). The dashed line indicates the chemical potential $\mu$. (b) Andreev bound states as a function of the superconducting phase mismatch $\phi$ for the Zeeman field $B_x=0.5\;\mathrm{meV}$ and $k_y=0$. The dashed lines indicate the topological phase transitions characterized by the topological invariant $\mathcal{Q}=\pm 1$ (framed insets). (c) Energy gap $E_g$ of the Andreev spectrum of the 2D Josephson junction at $k_y = 0$ as a function of $\phi$ and $B_x$. The topological regions in the phase diagram are separated by the gap closings (black lines). (d) Topological gap of the Andreev spectrum for all momenta [$E_g^{\text{top}}=\min_{k_y}E_g(k_y)$]. (e) - (h) Density of the lowest energy wavefunctions in real space for $\phi=1.2\pi$. (e) Non-localized Andreev bound state in the trivial region ($B_x = 0.05\; \mathrm{meV}$), (f) localized zero-energy Majorana bound state in the topological region ($B_x = 2.0\; \mathrm{meV}$), (g) and (h) two localized zero-energy states in the trivial region ($B_x = 4.0\; \mathrm{meV}$). (i) Conductance calculated at the edge of the N region of the JJ as a function of energy $E$ and $\phi$ at $B_x=0\mathrm{meV}$ and (j) $B_x=0.5\mathrm{meV}$ with a zero-bias peak in the topological region. Here, we used the parameters for Josephson junction based on $\alpha$-Sn $(\alpha_0,\alpha_z,\alpha_\square)=(18.62,11.88,0)\frac{\hbar^2}{2m}$ \cite{Madelung2004}, $\mu =-1.0\; \mathrm{meV}$, while the s-wave induced SC gap is taken for $\beta$-Sn with $\Delta=0.56\; \mathrm{meV}$ \cite{Ashcroft1976}. The dimensions of the junction are $W=20\;\mathrm{nm}$, $W_S=150\;\mathrm{nm}$, and $L=1000\;\mathrm{nm}$.}
    \label{fig:2}
\end{figure*}

In comparison with the 2-band Rashba model, the 4-band LM considers double degenerate LH and HH states. We show that the interplay of these two types of states in the LM gives rise to intriguing physics. 
The LM can describe both semimetal ($|\alpha_0|<|2\alpha_z - 3/5 \alpha_\square|$) and metal ($|\alpha_0|>|2\alpha_z - 3/5 \alpha_\square|$) regimes, where the LH and HH states have either opposite or the same curvature.

For the sake of simplicity, we study the bandstructure of the SC wire focusing on semimetals ($|\alpha_0|<|2\alpha_z - 3/5 \alpha_\square|$) and $\mu>0$, where the Fermi energy only lies in the LH band [Fig.~\ref{fig:1}(a)].
In the absence of a Zeeman field, the SC coupling $\Delta$ opens a gap in all states around the Fermi energy and we are in a trivial phase with $\mathcal{Q}=+1$.
Applying a Zeeman field $B_y$ breaks the time-reversal symmetry of the system. 
Consequently, the bands will split into positive and negative energies depending on their spin degeneracy. 
One can find a critical Zeeman field for each band where a topological phase transition occurs. This is caused by a band crossing at zero momentum. Such critical fields are given by
\begin{eqnarray}\label{Eq:Bcritical}
B_{\mathrm{HH}} = \frac{2}{3}\sqrt{\Delta^2 + \mu^2} \; , \; B_{\mathrm{LH}} = 2\sqrt{\Delta^2 + \mu^2}.
\end{eqnarray}
For $k=0$, the critical Zeeman field $B_y = B_{\mathrm{HH}}$ corresponds to a crossing between the quasi-electron and quasi-hole states in the BdG Hamiltonian for the $j_z = -3/2$ bands. 
This crossing yields a topological transition to a nontrivial phase characterized by a topological invariant $\mathcal{Q}=-1$ [Fig.~\ref{fig:1}(b)]. 
The second critical field $B_y = B_{\mathrm{LH}}$ denotes a band crossing between the quasi-electron and quasi-hole states for the $j_z = -1/2$ bands. 
In this regime, the system undergoes a second topological transition and returns to a trivial phase $\mathcal{Q} = +1$ [Fig.~\ref{fig:1}(c)]. The existence of two topological transitions can be used as an additional knob to identify MBS in experiments.
This last feature cannot be observed in 2-band wires with Rashba or Dresselhaus SOC \cite{Lutchyn2010,Oreg2010} since more than one topological transition can only occur in systems with more than two bands. 

We note that in the absence of BIA, the system is gapless [see Fig.~\ref{fig:1}(e)]. In 1D, the $\alpha_z$ and $\alpha_\square$ terms do not act as SOC and only give a difference in the effective masses of the $|j_z|=1/2$ and $|j_z|=3/2$ states (see Appendix). This is due to a simple fact that squared components of total angular and spin momenta give a constant.
Therefore, the gap cannot be opened with only the application of Zeeman fields and BIA has to be included as a source of SOC to find topological features in the system. 
High BIA strengths will enhance the topological gap. Hence, we expect the formation of MBSs in a wide range of materials from the $T_d$ symmetry class.

Due to the bulk-boundary correspondence \cite{Fukui2012}, the nontrivial topological invariant implies that we have a MBS localized at the edges of the wire. 
Using tight-biding calculations, we show in Fig.~\ref{fig:1}(f), that such zero-energy states indeed appear in the nontrivial region for $L\gg\lambda$, where $\lambda$ is the localization length of the MBS \cite{Prada2020}. As expected, the MBS at each end of a shorter wire hybridize and yield Majorana oscillations around zero energy. 

Similar results can be also obtained by considering Rashba SOC instead of BIA. 
However, we focuse here on BIA since it is an intrinsic property of the material, while Rashba SOC has to be generated by an applied electric field or asymmetric quantum well structure. 
We note that if Rashba SOC is considered in the 1D wire along the $x$-axis [Fig.~\ref{fig:setup}(a)], one requires a different direction of the Zeeman field ($B_x$) \cite{Lutchyn2010,Oreg2010}.

\section{\label{sec:2dJosephson} Two-dimensional Josephson junction}

In 2D Josephson junctions, the propagating particles in the N region experience multiple Andreev reflections at the NS interfaces giving rise to ABSs, that are confined to the N region [$|x|\leq W/2$, see Fig.\ref{fig:setup} (b)] \cite{Pientka2017,Scharf2019}. 
We first focus on the $k_y=0$ case of an infinite $L$ system, making the junction effectively 1D. 
In the absence of a Zeeman field, those ABSs are degenerate.
An in-plane Zeeman field applied to the N region breaks the time-reversal symmetry and lifts the degeneracy of the ABS [Fig.~\ref{fig:2}(b)]. 
This leads to two zero-energy crossings at critical phase differences, indicating an effective inversion of the gap and a change in the topology. Therefore, the energy gap as a function of Zeeman field and phase difference [see Fig.~\ref{fig:2}(c)] gives a topological phase diagram which can be alternatively obtained by calculating the topological invariant $\mathcal{Q}$. 
In a perfect system, without normal reflection at the NS-interface, the topological region is centered around the Thouless energy ($E_T= \frac{\pi}{2}\frac{\hbar v_F}{W}=2.2\;\mathrm{meV}$). 
Since the SC regions in our system have a finite length much larger than the coherence length \footnote{We estimate the coherence length in the SC regions to be $\xi=\frac{\hbar v_F}{\Delta}=50.0\;\mathrm{nm}$.}, we find a small contribution of normal reflection, which affects the topological phase diagram.
It was shown \cite{Kharitonov2017}, that any semi-infinite 2D Luttinger semimetal with a single edge and without cubic anisotropy ($\alpha_\square=0$) hosts either one (for $1 \leq |\alpha_0|/|\alpha_z| < 2$) or two (for $|\alpha_0|/|\alpha_z| < 1$) edge states, originating from the quadratic node in the bulk bandstructure. Metallic Luttinger materials ($|\alpha_0|/|\alpha_z| > 2$) do not host edge states. In this work, we focus on semimetals with one edge state at positive energies, using the material-specific parameters of $\alpha$-Sn ($\alpha_0/\alpha_z=1.57$) shown in Fig.~\ref{fig:2}(a). For such materials, the sign of $\mu$ is important and can be used to tune the system into different topological phases. In the subgap energy range ($|E| < |\Delta|$), the transmission in the N region is only given by the $j_z=\pm 3/2$ states for $\mu <0$ and determined by a combination of $j_z=\pm 1/2$ and edge states for $\mu>0$. From now on, we focus on the $\mu<0$ phase, without edge states.

Having a topological phase, which can host Majorana bound states, demands additionally a finite gap at all transverse momenta $k_y$. Interestingly, we find such gaps at all points in the topological region of Fig.~\ref{fig:2}(d) away from $\phi=0$, even without additional inversion symmetry breaking by BIA or Rashba terms. 
Previously, 2D Josephson junctions built from semiconductors were modeled by a 2DEG, where an additional inversion symmetry breaking Rashba or Dresselhaus term was needed to open a topological gap \cite{Hell2017,Pientka2017,Scharf2019,Ren2019}. 
In the Luttinger semimetal, the intrinsic SOC of the $\alpha_z$ and $\alpha_\square$ terms is $k_xk_y\{J_x,J_y\}$ (see Appendix). This term in combination with a finite phase difference between the SCs opens a topological gap with a nontrivial topological invariant under magnetic field. 

Considering a system that is confined in $y$-direction with a finite length $L$ allows us to study the wavefunctions of the states in real space. We show the density of the lowest-energy wavefunctions in the different regions of the topological phase diagram in Figs.~\ref{fig:2}(e)-(h) at $\phi=1.2\pi$. In the trivial region for small magnetic fields, we find ABS which are bound to the normal conducting region at $|x|\leq W/2$ [Fig.~\ref{fig:2}(e)] in the induced SC gap. For magnetic fields in the topological region with $\mathcal{Q}=-1$, we find a single zero-energy state which is additionally localized in $y$-direction giving rise to a MBS [Fig.~\ref{fig:2}(f)]. The localization length of the MBS is proportional to the inverse of the topological gap [$E_g^{\text{top}}(\phi,B_x)=\min_{k_y} E_g(\phi,B_x,k_y)$]. For increasing magnetic fields, the gap of the ABS closes again at $k_y=0$ and provides a second band inversion with the transition to trivial SC ($\mathcal{Q}=+1$). Here, we find two zero-energy states which are localized in $y$-direction [Figs.~\ref{fig:2}(g) and (h)] combining to a trivial conventional fermion.

Figs.~\ref{fig:2}(i) and (j) show calculations for the conductance of the system around the edge of the N region. Here, we added a small probe on top of the boundary of the N region, attached to a lead (see Fig.~\ref{fig:CondSetup}), similar as in Ref.~\cite{Ren2019}. Without magnetic field [Fig.~\ref{fig:2}(i)], one can clearly see the signal of the degenerate ABS¨ which cross at $\phi=\pi$. At finite magnetic field [Fig.~\ref{fig:2}(j)], the ABS split and host a topological region in between two crossings. The calculated conductance shows a clear zero bias peak in the topological region, induced by MBSs, which we predict to be observable in future experiments.

We emphasize, that the opening of a topological gap is independent of inversion symmetry breaking SOC terms, like BIA and Rashba, in a 2D Josephson junction on Luttinger semimetals. However, our findings are still applicable if such terms are present, as in tetrahedral materials, like HgTe. Therefore, we predict that any quadratic nodal Luttinger semimetal hosts a MBS in a 2D Josephson junction.

\section{\label{sec:conclusions} Conclusions and Outlook}
In this Letter, we analyze two different Luttinger semimetal systems that host topologically nontrivial properties leading to MBSs at zero energy. 
In SC wires, we find two topological transitions for two critical magnetic fields related to light an heavy hole band inversions. 
Interestingly, the range of magnetic fields where topological phase exists is material independent and only determined by the chemical potential. 
Moreover, we demonstrate that the intrinsic BIA term of any material of the tetrahedral symmetry group is sufficient for a gap opening in a SC wire, if the magnetic field is applied perpendicular to the wire.
This gap protects the formation of MBSs at zero energy.

In 2D Josephson junctions, we show that the intrinsic symmetric SOC of Luttinger materials in combination with the phase difference of the SCs is sufficient to generate MBSs, even without the application of BIA or Rashba SOC. 
The Luttinger model gives a more realistic description of typical semimetals, like $\alpha$-Sn or HgTe, than the usual 2DEGs with Rashba or Dresselhaus SOC. 
This opens a new avenue for the search of materials which should have intrinsically emergent Majorana bound states. 

Our results could also shed light on the formation of MBSs in other quadratic nodal semimetals, like Pr$_2$Ir$_2$O$_7$. Moreover, the competition between non-SC edge states [see Fig.~\ref{fig:2}(a)] in the Luttinger semimetal and the formation of MBSs could give rise to new physics that can be controlled by the chemical potential. Additionally, we expect interesting features in JJ on the metallic phase of $j=3/2$ Luttinger systems.

\section*{Acknowledgements}

We acknowledge funding by the Deutsche Forschungsgemeinschaft (DFG, German Research Foundation) through SFB 1170, Project-ID 258499086, through the Würzburg-Dresden Cluster of Excellence on Complexity and Topology in Quantum Matter –ct.qmat (EXC2147, Project-ID 390858490) as well as by the ENB Graduate School on Topological Insulators. We thank Ralph Claessen, Ion Cosma Fulga, Steffen Schreyeck and Jonas Erhardt for the enlightening discussions.

\appendix
\renewcommand{\theequation}{A\arabic{equation}}

\setcounter{equation}{0}

\renewcommand{\thefigure}{A\arabic{figure}}

\setcounter{figure}{0}

\section{Appendix A: Emergence of intrinsic spin-orbit coupling in 2D}

To study the effect of the symmetric SOC terms $\alpha_z$ and $\alpha_\square$, one can perform a basis rotation around the $y$-axis, via $\tilde{H}_L(\bm{k}) = \hat{D}_y^\dagger \hat{H}_L(\bm{k}) \hat{D}_y$. Here $\hat{D}_y = e^{-i \frac{\pi}{2}J_y}$ is the rotation operator which gives the Hamiltonian
\begin{widetext}
\begin{align}\label{eq:Halphaz}
 \tilde{H}_L(\bm{k}) = \begin{pmatrix}\alpha_0 \bm{k}^2 - \tilde{\alpha}(2k_x^2-k_y^2) -\mu & \sqrt{3}i \hat{\alpha} k_xk_y & \sqrt{3}\tilde{\alpha}k_y^2 & 0 \\ -\sqrt{3}i\hat{\alpha}k_xk_y & \alpha_0 \bm{k}^2 + \tilde{\alpha}(2k_x^2-k_y^2) - \mu & 0 & \sqrt{3} \tilde{\alpha}k_y^2 \\
 \sqrt{3}\tilde{\alpha} k_y^2 & 0 & \alpha_0 \bm{k}^2 + \tilde{\alpha}(2k_x^2-k_y^2) -\mu & - \sqrt{3}i\hat{\alpha} k_xk_y \\ 0 & \sqrt{3}\tilde{\alpha}k_y^2 & \sqrt{3}i\hat{\alpha}k_xk_y & \alpha_0 \bm{k}^2 - \tilde{\alpha}(2k_x^2-k_y^2)-\mu
\end{pmatrix},
\end{align}
\end{widetext}
with $\tilde{\alpha} = \alpha_z - \frac{3}{10}\alpha_\square$ and $\hat{\alpha} = 2\alpha_z + \frac{2}{5}\alpha_\square$. As one can see, in the 1D case for $k_y=0$, the Hamiltonian is purely diagonal. Therefore, $\alpha_z$ and $\alpha_\square$ do not couple the bands and do not act as SOC anymore.
In contrast, in 2D Luttinger materials, there is an off-diagonal term proportional to $k_xk_y$ which acts as an intrinsic SOC.

\section{Appendix B: Numerical conductance calculation}

\begin{figure}[h!]
 \centering
 \includegraphics[width=0.9\linewidth]{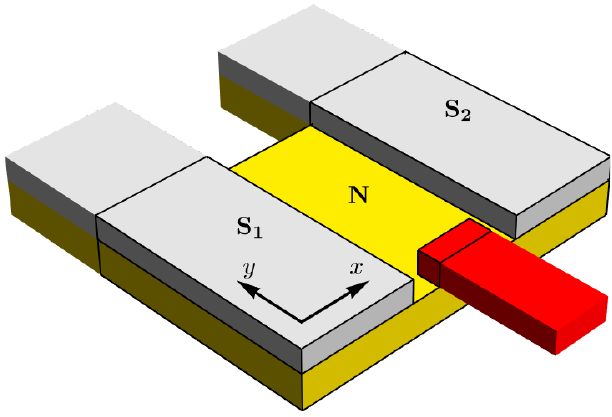}
 \caption{Setup of the Josephson junction modeled in the numerical calculations. To calculate the conductance of the edge states in the normal (N) region, we attach a probe on top of the edge of the N region (red) with a semi-infinite normal lead and attach two semi-infinite superconducting leads to the two superconducting regions ($S_1$, $S_2$). Finite size regions are indicated with a black border in the sketch, while semi-infinite leads are drawn without borders.}
 \label{fig:CondSetup}
\end{figure}

To calculate the conductance in the Josephson junction of Fig.~1(b), additional external leads are required. Since we are interested in the conductance of the Majorana bound states, which are localized at the boundaries in $y$-direction of the normal region, a normal conducting probe is attached as a second layer on top of this region and extends infinitely in $y\to - \infty$ (red part of Fig.~\ref{fig:CondSetup}). The probe is modeled with the normal Luttinger Hamiltonian $\hat{\tau_z}\hat{H}_L(\bm{k})$ of Eq.~(1). The probe is connected to the Josephson junction via a vertical tunnel coupling. The corresponding hopping term in the tight-binding Hamiltonian is given by 
\begin{align}
 H_{\text{probe coupling}}^{\text{TB}} = t_P \sum_{i=i_1}^{i_2} \sum_{j=j_1}^{j_2} d^\dagger_{i,j} \hat{\tau}_z \hat{1}_4 c_{i,j} + \text{h.c.}
\end{align}
Here, $i_1\leq i \leq i_2$ and $j_1\leq j \leq j_2$ indicate the $x$ and $y$ positions in the square lattice, corresponding to the contact area of $(10\times 25)\mathrm{nm}$. The creation (annihilation) operators $c_{i,j}^{\dagger}$ ($c_{i,j}$) act on the first layer of the Josephson junction and $d_{i,j}^{\dagger}$ ($d_{i,j}$) act on the second layer of the normal probe. For our calculations, we use the hopping strength $t_P= 0.5\mathrm{meV}$.
The two SC leads continue the SC regions of the Josephson junction in $y\to +\infty$ (grey part of Fig.~\ref{fig:CondSetup}).

The conductance between the normal lead and the two SC leads is determined, using the scattering matrix of the system at a given energy
\begin{align}
 G(E) = \frac{e^2}{h} \left[N - R_{ee}(E) + R_{eh}(E)\right],
\end{align}
where $N$ is the number of channels in the lead, and the reflection probabilities $R_{ee}(E)$ and $R_{eh}(E)$ correspond to normal and Andreev reflection in the normal lead.

\bibliographystyle{apsrev4-1}
\bibliography{LuttingerSC}

\begin{thebibliography}{53}%
\makeatletter
\providecommand \@ifxundefined [1]{%
 \@ifx{#1\undefined}
}%
\providecommand \@ifnum [1]{%
 \ifnum #1\expandafter \@firstoftwo
 \else \expandafter \@secondoftwo
 \fi
}%
\providecommand \@ifx [1]{%
 \ifx #1\expandafter \@firstoftwo
 \else \expandafter \@secondoftwo
 \fi
}%
\providecommand \natexlab [1]{#1}%
\providecommand \enquote  [1]{``#1''}%
\providecommand \bibnamefont  [1]{#1}%
\providecommand \bibfnamefont [1]{#1}%
\providecommand \citenamefont [1]{#1}%
\providecommand \href@noop [0]{\@secondoftwo}%
\providecommand \href [0]{\begingroup \@sanitize@url \@href}%
\providecommand \@href[1]{\@@startlink{#1}\@@href}%
\providecommand \@@href[1]{\endgroup#1\@@endlink}%
\providecommand \@sanitize@url [0]{\catcode `\\12\catcode `\$12\catcode
  `\&12\catcode `\#12\catcode `\^12\catcode `\_12\catcode `\%12\relax}%
\providecommand \@@startlink[1]{}%
\providecommand \@@endlink[0]{}%
\providecommand \url  [0]{\begingroup\@sanitize@url \@url }%
\providecommand \@url [1]{\endgroup\@href {#1}{\urlprefix }}%
\providecommand \urlprefix  [0]{URL }%
\providecommand \Eprint [0]{\href }%
\providecommand \doibase [0]{http://dx.doi.org/}%
\providecommand \selectlanguage [0]{\@gobble}%
\providecommand \bibinfo  [0]{\@secondoftwo}%
\providecommand \bibfield  [0]{\@secondoftwo}%
\providecommand \translation [1]{[#1]}%
\providecommand \BibitemOpen [0]{}%
\providecommand \bibitemStop [0]{}%
\providecommand \bibitemNoStop [0]{.\EOS\space}%
\providecommand \EOS [0]{\spacefactor3000\relax}%
\providecommand \BibitemShut  [1]{\csname bibitem#1\endcsname}%
\let\auto@bib@innerbib\@empty
\bibitem [{\citenamefont {Alicea}(2012)}]{Alicea2012}%
  \BibitemOpen
  \bibfield  {author} {\bibinfo {author} {\bibfnamefont {J.}~\bibnamefont
  {Alicea}},\ }\href {\doibase 10.1088/0034-4885/75/7/076501} {\bibfield
  {journal} {\bibinfo  {journal} {Reports on Progress in Physics}\ }\textbf
  {\bibinfo {volume} {75}},\ \bibinfo {pages} {076501} (\bibinfo {year}
  {2012})}\BibitemShut {NoStop}%
\bibitem [{\citenamefont {Beenakker}(2013)}]{Beenakker2013}%
  \BibitemOpen
  \bibfield  {author} {\bibinfo {author} {\bibfnamefont {C.}~\bibnamefont
  {Beenakker}},\ }\href {\doibase 10.1146/annurev-conmatphys-030212-184337}
  {\bibfield  {journal} {\bibinfo  {journal} {Annual Review of Condensed Matter
  Physics}\ }\textbf {\bibinfo {volume} {4}},\ \bibinfo {pages} {113} (\bibinfo
  {year} {2013})},\ \Eprint
  {http://arxiv.org/abs/https://doi.org/10.1146/annurev-conmatphys-030212-184337}
  {https://doi.org/10.1146/annurev-conmatphys-030212-184337} \BibitemShut
  {NoStop}%
\bibitem [{\citenamefont {Lutchyn}\ \emph {et~al.}(2018)\citenamefont
  {Lutchyn}, \citenamefont {Bakkers}, \citenamefont {Kouwenhoven},
  \citenamefont {Krogstrup}, \citenamefont {Marcus},\ and\ \citenamefont
  {Oreg}}]{Lutchyn2018}%
  \BibitemOpen
  \bibfield  {author} {\bibinfo {author} {\bibfnamefont {R.~M.}\ \bibnamefont
  {Lutchyn}}, \bibinfo {author} {\bibfnamefont {E.~P. A.~M.}\ \bibnamefont
  {Bakkers}}, \bibinfo {author} {\bibfnamefont {L.~P.}\ \bibnamefont
  {Kouwenhoven}}, \bibinfo {author} {\bibfnamefont {P.}~\bibnamefont
  {Krogstrup}}, \bibinfo {author} {\bibfnamefont {C.~M.}\ \bibnamefont
  {Marcus}}, \ and\ \bibinfo {author} {\bibfnamefont {Y.}~\bibnamefont
  {Oreg}},\ }\href {\doibase 10.1038/s41578-018-0003-1} {\bibfield  {journal}
  {\bibinfo  {journal} {Nature Reviews Materials}\ }\textbf {\bibinfo {volume}
  {3}},\ \bibinfo {pages} {52} (\bibinfo {year} {2018})}\BibitemShut {NoStop}%
\bibitem [{\citenamefont {Prada}\ \emph {et~al.}(2020)\citenamefont {Prada},
  \citenamefont {San-Jose}, \citenamefont {de~Moor}, \citenamefont {Geresdi},
  \citenamefont {Lee}, \citenamefont {Klinovaja}, \citenamefont {Loss},
  \citenamefont {Nyg{\aa}rd}, \citenamefont {Aguado},\ and\ \citenamefont
  {Kouwenhoven}}]{Prada2020}%
  \BibitemOpen
  \bibfield  {author} {\bibinfo {author} {\bibfnamefont {E.}~\bibnamefont
  {Prada}}, \bibinfo {author} {\bibfnamefont {P.}~\bibnamefont {San-Jose}},
  \bibinfo {author} {\bibfnamefont {M.~W.~A.}\ \bibnamefont {de~Moor}},
  \bibinfo {author} {\bibfnamefont {A.}~\bibnamefont {Geresdi}}, \bibinfo
  {author} {\bibfnamefont {E.~J.~H.}\ \bibnamefont {Lee}}, \bibinfo {author}
  {\bibfnamefont {J.}~\bibnamefont {Klinovaja}}, \bibinfo {author}
  {\bibfnamefont {D.}~\bibnamefont {Loss}}, \bibinfo {author} {\bibfnamefont
  {J.}~\bibnamefont {Nyg{\aa}rd}}, \bibinfo {author} {\bibfnamefont
  {R.}~\bibnamefont {Aguado}}, \ and\ \bibinfo {author} {\bibfnamefont {L.~P.}\
  \bibnamefont {Kouwenhoven}},\ }\href {\doibase 10.1038/s42254-020-0228-y}
  {\bibfield  {journal} {\bibinfo  {journal} {Nature Reviews Physics}\ }\textbf
  {\bibinfo {volume} {2}},\ \bibinfo {pages} {575} (\bibinfo {year} {2020})},\
  \Eprint {http://arxiv.org/abs/1911.04512} {arXiv:1911.04512} \BibitemShut
  {NoStop}%
\bibitem [{\citenamefont {Aguado}\ and\ \citenamefont
  {Kouwenhoven}(2020)}]{Aguado2020}%
  \BibitemOpen
  \bibfield  {author} {\bibinfo {author} {\bibfnamefont {R.}~\bibnamefont
  {Aguado}}\ and\ \bibinfo {author} {\bibfnamefont {L.~P.}\ \bibnamefont
  {Kouwenhoven}},\ }\href {\doibase 10.1063/PT.3.4499} {\bibfield  {journal}
  {\bibinfo  {journal} {Physics Today}\ }\textbf {\bibinfo {volume} {73}},\
  \bibinfo {pages} {44} (\bibinfo {year} {2020})}\BibitemShut {NoStop}%
\bibitem [{\citenamefont {Kitaev}(2003)}]{Kitaev2003}%
  \BibitemOpen
  \bibfield  {author} {\bibinfo {author} {\bibfnamefont {A.}~\bibnamefont
  {Kitaev}},\ }\href {\doibase 10.1016/S0003-4916(02)00018-0} {\bibfield
  {journal} {\bibinfo  {journal} {Annals of Physics}\ }\textbf {\bibinfo
  {volume} {303}},\ \bibinfo {pages} {2} (\bibinfo {year} {2003})}\BibitemShut
  {NoStop}%
\bibitem [{\citenamefont {Leijnse}\ and\ \citenamefont
  {Flensberg}(2012)}]{Leijnse2012}%
  \BibitemOpen
  \bibfield  {author} {\bibinfo {author} {\bibfnamefont {M.}~\bibnamefont
  {Leijnse}}\ and\ \bibinfo {author} {\bibfnamefont {K.}~\bibnamefont
  {Flensberg}},\ }\href {\doibase 10.1088/0268-1242/27/12/124003} {\bibfield
  {journal} {\bibinfo  {journal} {Semiconductor Science and Technology}\
  }\textbf {\bibinfo {volume} {27}},\ \bibinfo {pages} {124003} (\bibinfo
  {year} {2012})},\ \Eprint {http://arxiv.org/abs/1206.1736} {arXiv:1206.1736}
  \BibitemShut {NoStop}%
\bibitem [{\citenamefont {Fu}\ and\ \citenamefont {Kane}(2008)}]{Fu2008}%
  \BibitemOpen
  \bibfield  {author} {\bibinfo {author} {\bibfnamefont {L.}~\bibnamefont
  {Fu}}\ and\ \bibinfo {author} {\bibfnamefont {C.~L.}\ \bibnamefont {Kane}},\
  }\href {\doibase 10.1103/PhysRevLett.100.096407} {\bibfield  {journal}
  {\bibinfo  {journal} {Physical Review Letters}\ }\textbf {\bibinfo {volume}
  {100}},\ \bibinfo {pages} {096407} (\bibinfo {year} {2008})}\BibitemShut
  {NoStop}%
\bibitem [{\citenamefont {Goldstein}\ and\ \citenamefont
  {Chamon}(2011)}]{Goldstein2011}%
  \BibitemOpen
  \bibfield  {author} {\bibinfo {author} {\bibfnamefont {G.}~\bibnamefont
  {Goldstein}}\ and\ \bibinfo {author} {\bibfnamefont {C.}~\bibnamefont
  {Chamon}},\ }\href {\doibase 10.1103/PhysRevB.84.205109} {\bibfield
  {journal} {\bibinfo  {journal} {Physical Review B}\ }\textbf {\bibinfo
  {volume} {84}},\ \bibinfo {pages} {205109} (\bibinfo {year}
  {2011})}\BibitemShut {NoStop}%
\bibitem [{\citenamefont {Mourik}\ \emph {et~al.}(2012)\citenamefont {Mourik},
  \citenamefont {Zuo}, \citenamefont {Frolov}, \citenamefont {Plissard},
  \citenamefont {Bakkers},\ and\ \citenamefont {Kouwenhoven}}]{Mourik2012}%
  \BibitemOpen
  \bibfield  {author} {\bibinfo {author} {\bibfnamefont {V.}~\bibnamefont
  {Mourik}}, \bibinfo {author} {\bibfnamefont {K.}~\bibnamefont {Zuo}},
  \bibinfo {author} {\bibfnamefont {S.~M.}\ \bibnamefont {Frolov}}, \bibinfo
  {author} {\bibfnamefont {S.~R.}\ \bibnamefont {Plissard}}, \bibinfo {author}
  {\bibfnamefont {E.~P. A.~M.}\ \bibnamefont {Bakkers}}, \ and\ \bibinfo
  {author} {\bibfnamefont {L.~P.}\ \bibnamefont {Kouwenhoven}},\ }\href
  {\doibase 10.1126/science.1222360} {\bibfield  {journal} {\bibinfo  {journal}
  {Science}\ }\textbf {\bibinfo {volume} {336}},\ \bibinfo {pages} {1003}
  (\bibinfo {year} {2012})}\BibitemShut {NoStop}%
\bibitem [{\citenamefont {Deng}\ \emph {et~al.}(2018)\citenamefont {Deng},
  \citenamefont {Vaitiekėnas}, \citenamefont {Prada}, \citenamefont
  {San-Jose}, \citenamefont {Nyg{\aa}rd}, \citenamefont {Krogstrup},
  \citenamefont {Aguado},\ and\ \citenamefont {Marcus}}]{Deng2018}%
  \BibitemOpen
  \bibfield  {author} {\bibinfo {author} {\bibfnamefont {M.-T.}\ \bibnamefont
  {Deng}}, \bibinfo {author} {\bibfnamefont {S.}~\bibnamefont {Vaitiekėnas}},
  \bibinfo {author} {\bibfnamefont {E.}~\bibnamefont {Prada}}, \bibinfo
  {author} {\bibfnamefont {P.}~\bibnamefont {San-Jose}}, \bibinfo {author}
  {\bibfnamefont {J.}~\bibnamefont {Nyg{\aa}rd}}, \bibinfo {author}
  {\bibfnamefont {P.}~\bibnamefont {Krogstrup}}, \bibinfo {author}
  {\bibfnamefont {R.}~\bibnamefont {Aguado}}, \ and\ \bibinfo {author}
  {\bibfnamefont {C.~M.}\ \bibnamefont {Marcus}},\ }\href {\doibase
  10.1103/PhysRevB.98.085125} {\bibfield  {journal} {\bibinfo  {journal}
  {Physical Review B}\ }\textbf {\bibinfo {volume} {98}},\ \bibinfo {pages}
  {085125} (\bibinfo {year} {2018})},\ \Eprint
  {http://arxiv.org/abs/1712.03536} {arXiv:1712.03536} \BibitemShut {NoStop}%
\bibitem [{\citenamefont {Fornieri}\ \emph {et~al.}(2019)\citenamefont
  {Fornieri}, \citenamefont {Whiticar}, \citenamefont {Setiawan}, \citenamefont
  {Portol{\'{e}}s}, \citenamefont {Drachmann}, \citenamefont {Keselman},
  \citenamefont {Gronin}, \citenamefont {Thomas}, \citenamefont {Wang},
  \citenamefont {Kallaher}, \citenamefont {Gardner}, \citenamefont {Berg},
  \citenamefont {Manfra}, \citenamefont {Stern}, \citenamefont {Marcus},\ and\
  \citenamefont {Nichele}}]{Fornieri2019}%
  \BibitemOpen
  \bibfield  {author} {\bibinfo {author} {\bibfnamefont {A.}~\bibnamefont
  {Fornieri}}, \bibinfo {author} {\bibfnamefont {A.~M.}\ \bibnamefont
  {Whiticar}}, \bibinfo {author} {\bibfnamefont {F.}~\bibnamefont {Setiawan}},
  \bibinfo {author} {\bibfnamefont {E.}~\bibnamefont {Portol{\'{e}}s}},
  \bibinfo {author} {\bibfnamefont {A.~C.~C.}\ \bibnamefont {Drachmann}},
  \bibinfo {author} {\bibfnamefont {A.}~\bibnamefont {Keselman}}, \bibinfo
  {author} {\bibfnamefont {S.}~\bibnamefont {Gronin}}, \bibinfo {author}
  {\bibfnamefont {C.}~\bibnamefont {Thomas}}, \bibinfo {author} {\bibfnamefont
  {T.}~\bibnamefont {Wang}}, \bibinfo {author} {\bibfnamefont {R.}~\bibnamefont
  {Kallaher}}, \bibinfo {author} {\bibfnamefont {G.~C.}\ \bibnamefont
  {Gardner}}, \bibinfo {author} {\bibfnamefont {E.}~\bibnamefont {Berg}},
  \bibinfo {author} {\bibfnamefont {M.~J.}\ \bibnamefont {Manfra}}, \bibinfo
  {author} {\bibfnamefont {A.}~\bibnamefont {Stern}}, \bibinfo {author}
  {\bibfnamefont {C.~M.}\ \bibnamefont {Marcus}}, \ and\ \bibinfo {author}
  {\bibfnamefont {F.}~\bibnamefont {Nichele}},\ }\href {\doibase
  10.1038/s41586-019-1068-8} {\bibfield  {journal} {\bibinfo  {journal}
  {Nature}\ }\textbf {\bibinfo {volume} {569}},\ \bibinfo {pages} {89}
  (\bibinfo {year} {2019})},\ \Eprint {http://arxiv.org/abs/1809.03037}
  {arXiv:1809.03037} \BibitemShut {NoStop}%
\bibitem [{\citenamefont {Ren}\ \emph {et~al.}(2019)\citenamefont {Ren},
  \citenamefont {Pientka}, \citenamefont {Hart}, \citenamefont {Pierce},
  \citenamefont {Kosowsky}, \citenamefont {Lunczer}, \citenamefont {Schlereth},
  \citenamefont {Scharf}, \citenamefont {Hankiewicz}, \citenamefont
  {Molenkamp},\ and\ \citenamefont {et~al.}}]{Ren2019}%
  \BibitemOpen
  \bibfield  {author} {\bibinfo {author} {\bibfnamefont {H.}~\bibnamefont
  {Ren}}, \bibinfo {author} {\bibfnamefont {F.}~\bibnamefont {Pientka}},
  \bibinfo {author} {\bibfnamefont {S.}~\bibnamefont {Hart}}, \bibinfo {author}
  {\bibfnamefont {A.~T.}\ \bibnamefont {Pierce}}, \bibinfo {author}
  {\bibfnamefont {M.}~\bibnamefont {Kosowsky}}, \bibinfo {author}
  {\bibfnamefont {L.}~\bibnamefont {Lunczer}}, \bibinfo {author} {\bibfnamefont
  {R.}~\bibnamefont {Schlereth}}, \bibinfo {author} {\bibfnamefont
  {B.}~\bibnamefont {Scharf}}, \bibinfo {author} {\bibfnamefont {E.~M.}\
  \bibnamefont {Hankiewicz}}, \bibinfo {author} {\bibfnamefont {L.~W.}\
  \bibnamefont {Molenkamp}}, \ and\ \bibinfo {author} {\bibnamefont {et~al.}},\
  }\href {\doibase 10.1038/s41586-019-1148-9} {\bibfield  {journal} {\bibinfo
  {journal} {Nature}\ }\textbf {\bibinfo {volume} {569}},\ \bibinfo {pages}
  {93–98} (\bibinfo {year} {2019})}\BibitemShut {NoStop}%
\bibitem [{\citenamefont {Lutchyn}\ \emph {et~al.}(2010)\citenamefont
  {Lutchyn}, \citenamefont {Sau},\ and\ \citenamefont {{Das
  Sarma}}}]{Lutchyn2010}%
  \BibitemOpen
  \bibfield  {author} {\bibinfo {author} {\bibfnamefont {R.~M.}\ \bibnamefont
  {Lutchyn}}, \bibinfo {author} {\bibfnamefont {J.~D.}\ \bibnamefont {Sau}}, \
  and\ \bibinfo {author} {\bibfnamefont {S.}~\bibnamefont {{Das Sarma}}},\
  }\href {\doibase 10.1103/PhysRevLett.105.077001} {\bibfield  {journal}
  {\bibinfo  {journal} {Physical Review Letters}\ }\textbf {\bibinfo {volume}
  {105}},\ \bibinfo {pages} {077001} (\bibinfo {year} {2010})},\ \Eprint
  {http://arxiv.org/abs/1002.4033} {arXiv:1002.4033} \BibitemShut {NoStop}%
\bibitem [{\citenamefont {Oreg}\ \emph {et~al.}(2010)\citenamefont {Oreg},
  \citenamefont {Refael},\ and\ \citenamefont {von Oppen}}]{Oreg2010}%
  \BibitemOpen
  \bibfield  {author} {\bibinfo {author} {\bibfnamefont {Y.}~\bibnamefont
  {Oreg}}, \bibinfo {author} {\bibfnamefont {G.}~\bibnamefont {Refael}}, \ and\
  \bibinfo {author} {\bibfnamefont {F.}~\bibnamefont {von Oppen}},\ }\href
  {\doibase 10.1103/PhysRevLett.105.177002} {\bibfield  {journal} {\bibinfo
  {journal} {Physical Review Letters}\ }\textbf {\bibinfo {volume} {105}},\
  \bibinfo {pages} {177002} (\bibinfo {year} {2010})},\ \Eprint
  {http://arxiv.org/abs/1003.1145} {arXiv:1003.1145} \BibitemShut {NoStop}%
\bibitem [{\citenamefont {Wakatsuki}\ \emph {et~al.}(2014)\citenamefont
  {Wakatsuki}, \citenamefont {Ezawa},\ and\ \citenamefont
  {Nagaosa}}]{Wakatsuki2014}%
  \BibitemOpen
  \bibfield  {author} {\bibinfo {author} {\bibfnamefont {R.}~\bibnamefont
  {Wakatsuki}}, \bibinfo {author} {\bibfnamefont {M.}~\bibnamefont {Ezawa}}, \
  and\ \bibinfo {author} {\bibfnamefont {N.}~\bibnamefont {Nagaosa}},\ }\href
  {\doibase 10.1103/PhysRevB.89.174514} {\bibfield  {journal} {\bibinfo
  {journal} {Physical Review B}\ }\textbf {\bibinfo {volume} {89}},\ \bibinfo
  {pages} {174514} (\bibinfo {year} {2014})},\ \Eprint
  {http://arxiv.org/abs/1401.5192} {arXiv:1401.5192} \BibitemShut {NoStop}%
\bibitem [{\citenamefont {Szumniak}\ \emph {et~al.}(2017)\citenamefont
  {Szumniak}, \citenamefont {Chevallier}, \citenamefont {Loss},\ and\
  \citenamefont {Klinovaja}}]{Szumniak2017}%
  \BibitemOpen
  \bibfield  {author} {\bibinfo {author} {\bibfnamefont {P.}~\bibnamefont
  {Szumniak}}, \bibinfo {author} {\bibfnamefont {D.}~\bibnamefont
  {Chevallier}}, \bibinfo {author} {\bibfnamefont {D.}~\bibnamefont {Loss}}, \
  and\ \bibinfo {author} {\bibfnamefont {J.}~\bibnamefont {Klinovaja}},\ }\href
  {\doibase 10.1103/PhysRevB.96.041401} {\bibfield  {journal} {\bibinfo
  {journal} {Physical Review B}\ }\textbf {\bibinfo {volume} {96}},\ \bibinfo
  {pages} {041401} (\bibinfo {year} {2017})},\ \Eprint
  {http://arxiv.org/abs/1703.00265} {arXiv:1703.00265} \BibitemShut {NoStop}%
\bibitem [{\citenamefont {Kharitonov}\ \emph {et~al.}(2021)\citenamefont
  {Kharitonov}, \citenamefont {Hankiewicz}, \citenamefont {Trauzettel},\ and\
  \citenamefont {Bergeret}}]{Kharitonov2020}%
  \BibitemOpen
  \bibfield  {author} {\bibinfo {author} {\bibfnamefont {M.}~\bibnamefont
  {Kharitonov}}, \bibinfo {author} {\bibfnamefont {E.~M.}\ \bibnamefont
  {Hankiewicz}}, \bibinfo {author} {\bibfnamefont {B.}~\bibnamefont
  {Trauzettel}}, \ and\ \bibinfo {author} {\bibfnamefont {F.~S.}\ \bibnamefont
  {Bergeret}},\ }\href {\doibase 10.1103/PhysRevB.104.134516} {\bibfield
  {journal} {\bibinfo  {journal} {Phys. Rev. B}\ }\textbf {\bibinfo {volume}
  {104}},\ \bibinfo {pages} {134516} (\bibinfo {year} {2021})}\BibitemShut
  {NoStop}%
\bibitem [{\citenamefont {Peng}\ \emph {et~al.}(2016)\citenamefont {Peng},
  \citenamefont {Pientka}, \citenamefont {Berg}, \citenamefont {Oreg},\ and\
  \citenamefont {von Oppen}}]{Peng2016}%
  \BibitemOpen
  \bibfield  {author} {\bibinfo {author} {\bibfnamefont {Y.}~\bibnamefont
  {Peng}}, \bibinfo {author} {\bibfnamefont {F.}~\bibnamefont {Pientka}},
  \bibinfo {author} {\bibfnamefont {E.}~\bibnamefont {Berg}}, \bibinfo {author}
  {\bibfnamefont {Y.}~\bibnamefont {Oreg}}, \ and\ \bibinfo {author}
  {\bibfnamefont {F.}~\bibnamefont {von Oppen}},\ }\href {\doibase
  10.1103/PhysRevB.94.085409} {\bibfield  {journal} {\bibinfo  {journal}
  {Physical Review B}\ }\textbf {\bibinfo {volume} {94}},\ \bibinfo {pages}
  {085409} (\bibinfo {year} {2016})},\ \Eprint
  {http://arxiv.org/abs/1604.04287} {arXiv:1604.04287} \BibitemShut {NoStop}%
\bibitem [{\citenamefont {Pientka}\ \emph {et~al.}(2017)\citenamefont
  {Pientka}, \citenamefont {Keselman}, \citenamefont {Berg}, \citenamefont
  {Yacoby}, \citenamefont {Stern},\ and\ \citenamefont
  {Halperin}}]{Pientka2017}%
  \BibitemOpen
  \bibfield  {author} {\bibinfo {author} {\bibfnamefont {F.}~\bibnamefont
  {Pientka}}, \bibinfo {author} {\bibfnamefont {A.}~\bibnamefont {Keselman}},
  \bibinfo {author} {\bibfnamefont {E.}~\bibnamefont {Berg}}, \bibinfo {author}
  {\bibfnamefont {A.}~\bibnamefont {Yacoby}}, \bibinfo {author} {\bibfnamefont
  {A.}~\bibnamefont {Stern}}, \ and\ \bibinfo {author} {\bibfnamefont {B.~I.}\
  \bibnamefont {Halperin}},\ }\href {\doibase 10.1103/PhysRevX.7.021032}
  {\bibfield  {journal} {\bibinfo  {journal} {Physical Review X}\ }\textbf
  {\bibinfo {volume} {7}},\ \bibinfo {pages} {021032} (\bibinfo {year}
  {2017})},\ \Eprint {http://arxiv.org/abs/1609.09482} {arXiv:1609.09482}
  \BibitemShut {NoStop}%
\bibitem [{\citenamefont {Scharf}\ \emph {et~al.}(2019)\citenamefont {Scharf},
  \citenamefont {Pientka}, \citenamefont {Ren}, \citenamefont {Yacoby},\ and\
  \citenamefont {Hankiewicz}}]{Scharf2019}%
  \BibitemOpen
  \bibfield  {author} {\bibinfo {author} {\bibfnamefont {B.}~\bibnamefont
  {Scharf}}, \bibinfo {author} {\bibfnamefont {F.}~\bibnamefont {Pientka}},
  \bibinfo {author} {\bibfnamefont {H.}~\bibnamefont {Ren}}, \bibinfo {author}
  {\bibfnamefont {A.}~\bibnamefont {Yacoby}}, \ and\ \bibinfo {author}
  {\bibfnamefont {E.~M.}\ \bibnamefont {Hankiewicz}},\ }\href {\doibase
  10.1103/PhysRevB.99.214503} {\bibfield  {journal} {\bibinfo  {journal}
  {Physical Review B}\ }\textbf {\bibinfo {volume} {99}},\ \bibinfo {pages}
  {214503} (\bibinfo {year} {2019})},\ \Eprint
  {http://arxiv.org/abs/1904.08981} {arXiv:1904.08981} \BibitemShut {NoStop}%
\bibitem [{\citenamefont {Ruan}\ \emph {et~al.}(2016)\citenamefont {Ruan},
  \citenamefont {Jian}, \citenamefont {Yao}, \citenamefont {Zhang},
  \citenamefont {Zhang},\ and\ \citenamefont {Xing}}]{Ruan2016}%
  \BibitemOpen
  \bibfield  {author} {\bibinfo {author} {\bibfnamefont {J.}~\bibnamefont
  {Ruan}}, \bibinfo {author} {\bibfnamefont {S.-K.}\ \bibnamefont {Jian}},
  \bibinfo {author} {\bibfnamefont {H.}~\bibnamefont {Yao}}, \bibinfo {author}
  {\bibfnamefont {H.}~\bibnamefont {Zhang}}, \bibinfo {author} {\bibfnamefont
  {S.-C.}\ \bibnamefont {Zhang}}, \ and\ \bibinfo {author} {\bibfnamefont
  {D.}~\bibnamefont {Xing}},\ }\href {\doibase 10.1038/ncomms11136} {\bibfield
  {journal} {\bibinfo  {journal} {Nature Communications}\ }\textbf {\bibinfo
  {volume} {7}},\ \bibinfo {pages} {11136} (\bibinfo {year} {2016})},\ \Eprint
  {http://arxiv.org/abs/1511.08284} {arXiv:1511.08284} \BibitemShut {NoStop}%
\bibitem [{\citenamefont {Kharitonov}\ \emph {et~al.}(2017)\citenamefont
  {Kharitonov}, \citenamefont {Mayer},\ and\ \citenamefont
  {Hankiewicz}}]{Kharitonov2017}%
  \BibitemOpen
  \bibfield  {author} {\bibinfo {author} {\bibfnamefont {M.}~\bibnamefont
  {Kharitonov}}, \bibinfo {author} {\bibfnamefont {J.-B.}\ \bibnamefont
  {Mayer}}, \ and\ \bibinfo {author} {\bibfnamefont {E.~M.}\ \bibnamefont
  {Hankiewicz}},\ }\href {\doibase 10.1103/PhysRevLett.119.266402} {\bibfield
  {journal} {\bibinfo  {journal} {Physical Review Letters}\ }\textbf {\bibinfo
  {volume} {119}},\ \bibinfo {pages} {266402} (\bibinfo {year}
  {2017})}\BibitemShut {NoStop}%
\bibitem [{\citenamefont {Br{\"{u}}ne}\ \emph {et~al.}(2011)\citenamefont
  {Br{\"{u}}ne}, \citenamefont {Liu}, \citenamefont {Novik}, \citenamefont
  {Hankiewicz}, \citenamefont {Buhmann}, \citenamefont {Chen}, \citenamefont
  {Qi}, \citenamefont {Shen}, \citenamefont {Zhang},\ and\ \citenamefont
  {Molenkamp}}]{Brune2011}%
  \BibitemOpen
  \bibfield  {author} {\bibinfo {author} {\bibfnamefont {C.}~\bibnamefont
  {Br{\"{u}}ne}}, \bibinfo {author} {\bibfnamefont {C.~X.}\ \bibnamefont
  {Liu}}, \bibinfo {author} {\bibfnamefont {E.~G.}\ \bibnamefont {Novik}},
  \bibinfo {author} {\bibfnamefont {E.~M.}\ \bibnamefont {Hankiewicz}},
  \bibinfo {author} {\bibfnamefont {H.}~\bibnamefont {Buhmann}}, \bibinfo
  {author} {\bibfnamefont {Y.~L.}\ \bibnamefont {Chen}}, \bibinfo {author}
  {\bibfnamefont {X.~L.}\ \bibnamefont {Qi}}, \bibinfo {author} {\bibfnamefont
  {Z.~X.}\ \bibnamefont {Shen}}, \bibinfo {author} {\bibfnamefont {S.~C.}\
  \bibnamefont {Zhang}}, \ and\ \bibinfo {author} {\bibfnamefont {L.~W.}\
  \bibnamefont {Molenkamp}},\ }\href {\doibase 10.1103/PhysRevLett.106.126803}
  {\bibfield  {journal} {\bibinfo  {journal} {Physical Review Letters}\
  }\textbf {\bibinfo {volume} {106}},\ \bibinfo {pages} {126803} (\bibinfo
  {year} {2011})},\ \Eprint {http://arxiv.org/abs/1101.2627} {arXiv:1101.2627}
  \BibitemShut {NoStop}%
\bibitem [{\citenamefont {Br{\"{u}}ne}\ \emph {et~al.}(2014)\citenamefont
  {Br{\"{u}}ne}, \citenamefont {Thienel}, \citenamefont {Stuiber},
  \citenamefont {B{\"{o}}ttcher}, \citenamefont {Buhmann}, \citenamefont
  {Novik}, \citenamefont {Liu}, \citenamefont {Hankiewicz},\ and\ \citenamefont
  {Molenkamp}}]{Brune2014}%
  \BibitemOpen
  \bibfield  {author} {\bibinfo {author} {\bibfnamefont {C.}~\bibnamefont
  {Br{\"{u}}ne}}, \bibinfo {author} {\bibfnamefont {C.}~\bibnamefont
  {Thienel}}, \bibinfo {author} {\bibfnamefont {M.}~\bibnamefont {Stuiber}},
  \bibinfo {author} {\bibfnamefont {J.}~\bibnamefont {B{\"{o}}ttcher}},
  \bibinfo {author} {\bibfnamefont {H.}~\bibnamefont {Buhmann}}, \bibinfo
  {author} {\bibfnamefont {E.~G.}\ \bibnamefont {Novik}}, \bibinfo {author}
  {\bibfnamefont {C.-X.}\ \bibnamefont {Liu}}, \bibinfo {author} {\bibfnamefont
  {E.~M.}\ \bibnamefont {Hankiewicz}}, \ and\ \bibinfo {author} {\bibfnamefont
  {L.~W.}\ \bibnamefont {Molenkamp}},\ }\href {\doibase
  10.1103/PhysRevX.4.041045} {\bibfield  {journal} {\bibinfo  {journal}
  {Physical Review X}\ }\textbf {\bibinfo {volume} {4}},\ \bibinfo {pages}
  {041045} (\bibinfo {year} {2014})},\ \Eprint {http://arxiv.org/abs/1407.6537}
  {arXiv:1407.6537} \BibitemShut {NoStop}%
\bibitem [{\citenamefont {Baum}\ \emph {et~al.}(2014)\citenamefont {Baum},
  \citenamefont {B{\"{o}}ttcher}, \citenamefont {Br{\"{u}}ne}, \citenamefont
  {Thienel}, \citenamefont {Molenkamp}, \citenamefont {Stern},\ and\
  \citenamefont {Hankiewicz}}]{Baum2014}%
  \BibitemOpen
  \bibfield  {author} {\bibinfo {author} {\bibfnamefont {Y.}~\bibnamefont
  {Baum}}, \bibinfo {author} {\bibfnamefont {J.}~\bibnamefont
  {B{\"{o}}ttcher}}, \bibinfo {author} {\bibfnamefont {C.}~\bibnamefont
  {Br{\"{u}}ne}}, \bibinfo {author} {\bibfnamefont {C.}~\bibnamefont
  {Thienel}}, \bibinfo {author} {\bibfnamefont {L.~W.}\ \bibnamefont
  {Molenkamp}}, \bibinfo {author} {\bibfnamefont {A.}~\bibnamefont {Stern}}, \
  and\ \bibinfo {author} {\bibfnamefont {E.~M.}\ \bibnamefont {Hankiewicz}},\
  }\href {\doibase 10.1103/PhysRevB.89.245136} {\bibfield  {journal} {\bibinfo
  {journal} {Physical Review B}\ }\textbf {\bibinfo {volume} {89}},\ \bibinfo
  {pages} {245136} (\bibinfo {year} {2014})}\BibitemShut {NoStop}%
\bibitem [{\citenamefont {Morimoto}\ and\ \citenamefont
  {Furusaki}(2014)}]{Morimoto2014}%
  \BibitemOpen
  \bibfield  {author} {\bibinfo {author} {\bibfnamefont {T.}~\bibnamefont
  {Morimoto}}\ and\ \bibinfo {author} {\bibfnamefont {A.}~\bibnamefont
  {Furusaki}},\ }\href {\doibase 10.1103/PhysRevB.89.235127} {\bibfield
  {journal} {\bibinfo  {journal} {Physical Review B}\ }\textbf {\bibinfo
  {volume} {89}},\ \bibinfo {pages} {235127} (\bibinfo {year} {2014})},\
  \Eprint {http://arxiv.org/abs/1403.7962} {arXiv:1403.7962} \BibitemShut
  {NoStop}%
\bibitem [{\citenamefont {Kondo}\ \emph {et~al.}(2015)\citenamefont {Kondo},
  \citenamefont {Nakayama}, \citenamefont {Chen}, \citenamefont {Ishikawa},
  \citenamefont {Moon}, \citenamefont {Yamamoto}, \citenamefont {Ota},
  \citenamefont {Malaeb}, \citenamefont {Kanai}, \citenamefont {Nakashima},
  \citenamefont {Ishida}, \citenamefont {Yoshida}, \citenamefont {Yamamoto},
  \citenamefont {Matsunami}, \citenamefont {Kimura}, \citenamefont {Inami},
  \citenamefont {Ono}, \citenamefont {Kumigashira}, \citenamefont {Nakatsuji},
  \citenamefont {Balents},\ and\ \citenamefont {Shin}}]{Kondo2015}%
  \BibitemOpen
  \bibfield  {author} {\bibinfo {author} {\bibfnamefont {T.}~\bibnamefont
  {Kondo}}, \bibinfo {author} {\bibfnamefont {M.}~\bibnamefont {Nakayama}},
  \bibinfo {author} {\bibfnamefont {R.}~\bibnamefont {Chen}}, \bibinfo {author}
  {\bibfnamefont {J.~J.}\ \bibnamefont {Ishikawa}}, \bibinfo {author}
  {\bibfnamefont {E.-G.}\ \bibnamefont {Moon}}, \bibinfo {author}
  {\bibfnamefont {T.}~\bibnamefont {Yamamoto}}, \bibinfo {author}
  {\bibfnamefont {Y.}~\bibnamefont {Ota}}, \bibinfo {author} {\bibfnamefont
  {W.}~\bibnamefont {Malaeb}}, \bibinfo {author} {\bibfnamefont
  {H.}~\bibnamefont {Kanai}}, \bibinfo {author} {\bibfnamefont
  {Y.}~\bibnamefont {Nakashima}}, \bibinfo {author} {\bibfnamefont
  {Y.}~\bibnamefont {Ishida}}, \bibinfo {author} {\bibfnamefont
  {R.}~\bibnamefont {Yoshida}}, \bibinfo {author} {\bibfnamefont
  {H.}~\bibnamefont {Yamamoto}}, \bibinfo {author} {\bibfnamefont
  {M.}~\bibnamefont {Matsunami}}, \bibinfo {author} {\bibfnamefont
  {S.}~\bibnamefont {Kimura}}, \bibinfo {author} {\bibfnamefont
  {N.}~\bibnamefont {Inami}}, \bibinfo {author} {\bibfnamefont
  {K.}~\bibnamefont {Ono}}, \bibinfo {author} {\bibfnamefont {H.}~\bibnamefont
  {Kumigashira}}, \bibinfo {author} {\bibfnamefont {S.}~\bibnamefont
  {Nakatsuji}}, \bibinfo {author} {\bibfnamefont {L.}~\bibnamefont {Balents}},
  \ and\ \bibinfo {author} {\bibfnamefont {S.}~\bibnamefont {Shin}},\ }\href
  {\doibase 10.1038/ncomms10042} {\bibfield  {journal} {\bibinfo  {journal}
  {Nature Communications}\ }\textbf {\bibinfo {volume} {6}},\ \bibinfo {pages}
  {10042} (\bibinfo {year} {2015})},\ \Eprint {http://arxiv.org/abs/1510.07977}
  {arXiv:1510.07977} \BibitemShut {NoStop}%
\bibitem [{\citenamefont {Zhang}\ \emph {et~al.}(2018)\citenamefont {Zhang},
  \citenamefont {Wang}, \citenamefont {Ruan}, \citenamefont {Yao},\ and\
  \citenamefont {Zhang}}]{Zhang2018}%
  \BibitemOpen
  \bibfield  {author} {\bibinfo {author} {\bibfnamefont {D.}~\bibnamefont
  {Zhang}}, \bibinfo {author} {\bibfnamefont {H.}~\bibnamefont {Wang}},
  \bibinfo {author} {\bibfnamefont {J.}~\bibnamefont {Ruan}}, \bibinfo {author}
  {\bibfnamefont {G.}~\bibnamefont {Yao}}, \ and\ \bibinfo {author}
  {\bibfnamefont {H.}~\bibnamefont {Zhang}},\ }\href {\doibase
  10.1103/PhysRevB.97.195139} {\bibfield  {journal} {\bibinfo  {journal}
  {Physical Review B}\ }\textbf {\bibinfo {volume} {97}},\ \bibinfo {pages}
  {195139} (\bibinfo {year} {2018})},\ \Eprint
  {http://arxiv.org/abs/1802.01927} {arXiv:1802.01927} \BibitemShut {NoStop}%
\bibitem [{\citenamefont {Ghorashi}\ \emph {et~al.}(2018)\citenamefont
  {Ghorashi}, \citenamefont {Hosur},\ and\ \citenamefont
  {Ting}}]{Ghorashi2018}%
  \BibitemOpen
  \bibfield  {author} {\bibinfo {author} {\bibfnamefont {S.~A.~A.}\
  \bibnamefont {Ghorashi}}, \bibinfo {author} {\bibfnamefont {P.}~\bibnamefont
  {Hosur}}, \ and\ \bibinfo {author} {\bibfnamefont {C.-S.}\ \bibnamefont
  {Ting}},\ }\href {\doibase 10.1103/PhysRevB.97.205402} {\bibfield  {journal}
  {\bibinfo  {journal} {Physical Review B}\ }\textbf {\bibinfo {volume} {97}},\
  \bibinfo {pages} {205402} (\bibinfo {year} {2018})},\ \Eprint
  {http://arxiv.org/abs/1801.04287} {arXiv:1801.04287} \BibitemShut {NoStop}%
\bibitem [{\citenamefont {Boettcher}\ and\ \citenamefont
  {Herbut}(2016)}]{Boettcher2016}%
  \BibitemOpen
  \bibfield  {author} {\bibinfo {author} {\bibfnamefont {I.}~\bibnamefont
  {Boettcher}}\ and\ \bibinfo {author} {\bibfnamefont {I.~F.}\ \bibnamefont
  {Herbut}},\ }\href {\doibase 10.1103/PhysRevB.93.205138} {\bibfield
  {journal} {\bibinfo  {journal} {Physical Review B}\ }\textbf {\bibinfo
  {volume} {93}},\ \bibinfo {pages} {205138} (\bibinfo {year} {2016})},\
  \Eprint {http://arxiv.org/abs/1603.00031} {arXiv:1603.00031} \BibitemShut
  {NoStop}%
\bibitem [{\citenamefont {Boettcher}\ and\ \citenamefont
  {Herbut}(2018)}]{Boettcher2018}%
  \BibitemOpen
  \bibfield  {author} {\bibinfo {author} {\bibfnamefont {I.}~\bibnamefont
  {Boettcher}}\ and\ \bibinfo {author} {\bibfnamefont {I.~F.}\ \bibnamefont
  {Herbut}},\ }\href {\doibase 10.1103/PhysRevLett.120.057002} {\bibfield
  {journal} {\bibinfo  {journal} {Physical Review Letters}\ }\textbf {\bibinfo
  {volume} {120}},\ \bibinfo {pages} {057002} (\bibinfo {year} {2018})},\
  \Eprint {http://arxiv.org/abs/1707.03444} {arXiv:1707.03444} \BibitemShut
  {NoStop}%
\bibitem [{\citenamefont {Brydon}\ \emph {et~al.}(2016)\citenamefont {Brydon},
  \citenamefont {Wang}, \citenamefont {Weinert},\ and\ \citenamefont
  {Agterberg}}]{Brydon2016}%
  \BibitemOpen
  \bibfield  {author} {\bibinfo {author} {\bibfnamefont {P.}~\bibnamefont
  {Brydon}}, \bibinfo {author} {\bibfnamefont {L.}~\bibnamefont {Wang}},
  \bibinfo {author} {\bibfnamefont {M.}~\bibnamefont {Weinert}}, \ and\
  \bibinfo {author} {\bibfnamefont {D.}~\bibnamefont {Agterberg}},\ }\href
  {\doibase 10.1103/physrevlett.116.177001} {\bibfield  {journal} {\bibinfo
  {journal} {Physical Review Letters}\ }\textbf {\bibinfo {volume} {116}}
  (\bibinfo {year} {2016}),\ 10.1103/physrevlett.116.177001}\BibitemShut
  {NoStop}%
\bibitem [{\citenamefont {Kim}\ \emph {et~al.}(2018)\citenamefont {Kim},
  \citenamefont {Wang}, \citenamefont {Nakajima}, \citenamefont {Hu},
  \citenamefont {Ziemak}, \citenamefont {Syers}, \citenamefont {Wang},
  \citenamefont {Hodovanets}, \citenamefont {Denlinger}, \citenamefont
  {Brydon}, \citenamefont {Agterberg}, \citenamefont {Tanatar}, \citenamefont
  {Prozorov},\ and\ \citenamefont {Paglione}}]{Kim2018}%
  \BibitemOpen
  \bibfield  {author} {\bibinfo {author} {\bibfnamefont {H.}~\bibnamefont
  {Kim}}, \bibinfo {author} {\bibfnamefont {K.}~\bibnamefont {Wang}}, \bibinfo
  {author} {\bibfnamefont {Y.}~\bibnamefont {Nakajima}}, \bibinfo {author}
  {\bibfnamefont {R.}~\bibnamefont {Hu}}, \bibinfo {author} {\bibfnamefont
  {S.}~\bibnamefont {Ziemak}}, \bibinfo {author} {\bibfnamefont
  {P.}~\bibnamefont {Syers}}, \bibinfo {author} {\bibfnamefont
  {L.}~\bibnamefont {Wang}}, \bibinfo {author} {\bibfnamefont {H.}~\bibnamefont
  {Hodovanets}}, \bibinfo {author} {\bibfnamefont {J.~D.}\ \bibnamefont
  {Denlinger}}, \bibinfo {author} {\bibfnamefont {P.~M.~R.}\ \bibnamefont
  {Brydon}}, \bibinfo {author} {\bibfnamefont {D.~F.}\ \bibnamefont
  {Agterberg}}, \bibinfo {author} {\bibfnamefont {M.~A.}\ \bibnamefont
  {Tanatar}}, \bibinfo {author} {\bibfnamefont {R.}~\bibnamefont {Prozorov}}, \
  and\ \bibinfo {author} {\bibfnamefont {J.}~\bibnamefont {Paglione}},\ }\href
  {\doibase 10.1126/sciadv.aao4513} {\bibfield  {journal} {\bibinfo  {journal}
  {Science Advances}\ }\textbf {\bibinfo {volume} {4}},\ \bibinfo {pages}
  {eaao4513} (\bibinfo {year} {2018})},\ \Eprint
  {http://arxiv.org/abs/https://www.science.org/doi/pdf/10.1126/sciadv.aao4513}
  {https://www.science.org/doi/pdf/10.1126/sciadv.aao4513} \BibitemShut
  {NoStop}%
\bibitem [{\citenamefont {Roy}\ \emph {et~al.}(2019)\citenamefont {Roy},
  \citenamefont {Ghorashi}, \citenamefont {Foster},\ and\ \citenamefont
  {Nevidomskyy}}]{Roy2019}%
  \BibitemOpen
  \bibfield  {author} {\bibinfo {author} {\bibfnamefont {B.}~\bibnamefont
  {Roy}}, \bibinfo {author} {\bibfnamefont {S.~A.~A.}\ \bibnamefont
  {Ghorashi}}, \bibinfo {author} {\bibfnamefont {M.~S.}\ \bibnamefont
  {Foster}}, \ and\ \bibinfo {author} {\bibfnamefont {A.~H.}\ \bibnamefont
  {Nevidomskyy}},\ }\href {\doibase 10.1103/PhysRevB.99.054505} {\bibfield
  {journal} {\bibinfo  {journal} {Physical Review B}\ }\textbf {\bibinfo
  {volume} {99}},\ \bibinfo {pages} {054505} (\bibinfo {year} {2019})},\
  \Eprint {http://arxiv.org/abs/1708.07825} {arXiv:1708.07825} \BibitemShut
  {NoStop}%
\bibitem [{\citenamefont {Kim}\ \emph {et~al.}(2020)\citenamefont {Kim},
  \citenamefont {Lee}, \citenamefont {Hodovanets}, \citenamefont {Wang},
  \citenamefont {Sau},\ and\ \citenamefont {Paglione}}]{Kim2020}%
  \BibitemOpen
  \bibfield  {author} {\bibinfo {author} {\bibfnamefont {H.}~\bibnamefont
  {Kim}}, \bibinfo {author} {\bibfnamefont {J.}~\bibnamefont {Lee}}, \bibinfo
  {author} {\bibfnamefont {H.}~\bibnamefont {Hodovanets}}, \bibinfo {author}
  {\bibfnamefont {K.}~\bibnamefont {Wang}}, \bibinfo {author} {\bibfnamefont
  {J.~D.}\ \bibnamefont {Sau}}, \ and\ \bibinfo {author} {\bibfnamefont
  {J.}~\bibnamefont {Paglione}},\ }\href {https://arxiv.org/abs/2010.12085} {\
  (\bibinfo {year} {2020})},\ \Eprint {http://arxiv.org/abs/2010.12085}
  {arXiv:2010.12085} \BibitemShut {NoStop}%
\bibitem [{\citenamefont {Dutta}\ \emph {et~al.}(2021)\citenamefont {Dutta},
  \citenamefont {Parhizgar},\ and\ \citenamefont {Black-Schaffer}}]{Dutta2021}%
  \BibitemOpen
  \bibfield  {author} {\bibinfo {author} {\bibfnamefont {P.}~\bibnamefont
  {Dutta}}, \bibinfo {author} {\bibfnamefont {F.}~\bibnamefont {Parhizgar}}, \
  and\ \bibinfo {author} {\bibfnamefont {A.~M.}\ \bibnamefont
  {Black-Schaffer}},\ }\href {\doibase 10.1103/PhysRevResearch.3.033255}
  {\bibfield  {journal} {\bibinfo  {journal} {Phys. Rev. Research}\ }\textbf
  {\bibinfo {volume} {3}},\ \bibinfo {pages} {033255} (\bibinfo {year}
  {2021})}\BibitemShut {NoStop}%
\bibitem [{\citenamefont {Timm}\ and\ \citenamefont
  {Bhattacharya}(2021)}]{Timm2021}%
  \BibitemOpen
  \bibfield  {author} {\bibinfo {author} {\bibfnamefont {C.}~\bibnamefont
  {Timm}}\ and\ \bibinfo {author} {\bibfnamefont {A.}~\bibnamefont
  {Bhattacharya}},\ }\href {\doibase 10.1103/PhysRevB.104.094529} {\bibfield
  {journal} {\bibinfo  {journal} {Phys. Rev. B}\ }\textbf {\bibinfo {volume}
  {104}},\ \bibinfo {pages} {094529} (\bibinfo {year} {2021})}\BibitemShut
  {NoStop}%
\bibitem [{\citenamefont {Bahari}\ \emph {et~al.}(2022)\citenamefont {Bahari},
  \citenamefont {Zhang},\ and\ \citenamefont {Trauzettel}}]{Bahari2022}%
  \BibitemOpen
  \bibfield  {author} {\bibinfo {author} {\bibfnamefont {M.}~\bibnamefont
  {Bahari}}, \bibinfo {author} {\bibfnamefont {S.-B.}\ \bibnamefont {Zhang}}, \
  and\ \bibinfo {author} {\bibfnamefont {B.}~\bibnamefont {Trauzettel}},\
  }\href {\doibase 10.1103/PhysRevResearch.4.L012017} {\bibfield  {journal}
  {\bibinfo  {journal} {Phys. Rev. Research}\ }\textbf {\bibinfo {volume}
  {4}},\ \bibinfo {pages} {L012017} (\bibinfo {year} {2022})}\BibitemShut
  {NoStop}%
\bibitem [{\citenamefont {Winkler}(2003)}]{Winkler2003}%
  \BibitemOpen
  \bibfield  {author} {\bibinfo {author} {\bibfnamefont {R.}~\bibnamefont
  {Winkler}},\ }\href {\doibase 10.1007/b13586} {\emph {\bibinfo {title}
  {{Spin--Orbit Coupling Effects in Two-Dimensional Electron and Hole
  Systems}}}},\ \bibinfo {series} {Springer Tracts in Modern Physics}, Vol.\
  \bibinfo {volume} {191}\ (\bibinfo  {publisher} {Springer Berlin
  Heidelberg},\ \bibinfo {address} {Berlin, Heidelberg},\ \bibinfo {year}
  {2003})\BibitemShut {NoStop}%
\bibitem [{\citenamefont {Kohn}\ and\ \citenamefont
  {Luttinger}(1954)}]{Kohn1954}%
  \BibitemOpen
  \bibfield  {author} {\bibinfo {author} {\bibfnamefont {W.}~\bibnamefont
  {Kohn}}\ and\ \bibinfo {author} {\bibfnamefont {J.~M.}\ \bibnamefont
  {Luttinger}},\ }\href {\doibase 10.1103/PhysRev.96.529.2} {\bibfield
  {journal} {\bibinfo  {journal} {Physical Review}\ }\textbf {\bibinfo {volume}
  {96}},\ \bibinfo {pages} {529} (\bibinfo {year} {1954})}\BibitemShut
  {NoStop}%
\bibitem [{\citenamefont {Schnyder}\ \emph {et~al.}(2008)\citenamefont
  {Schnyder}, \citenamefont {Ryu}, \citenamefont {Furusaki},\ and\
  \citenamefont {Ludwig}}]{Schnyder2008}%
  \BibitemOpen
  \bibfield  {author} {\bibinfo {author} {\bibfnamefont {A.~P.}\ \bibnamefont
  {Schnyder}}, \bibinfo {author} {\bibfnamefont {S.}~\bibnamefont {Ryu}},
  \bibinfo {author} {\bibfnamefont {A.}~\bibnamefont {Furusaki}}, \ and\
  \bibinfo {author} {\bibfnamefont {A.~W.~W.}\ \bibnamefont {Ludwig}},\ }\href
  {\doibase 10.1103/PhysRevB.78.195125} {\bibfield  {journal} {\bibinfo
  {journal} {Physical Review B}\ }\textbf {\bibinfo {volume} {78}},\ \bibinfo
  {pages} {195125} (\bibinfo {year} {2008})},\ \Eprint
  {http://arxiv.org/abs/0803.2786} {arXiv:0803.2786} \BibitemShut {NoStop}%
\bibitem [{\citenamefont {Ryu}\ \emph {et~al.}(2010)\citenamefont {Ryu},
  \citenamefont {Schnyder}, \citenamefont {Furusaki},\ and\ \citenamefont
  {Ludwig}}]{Ryu2010}%
  \BibitemOpen
  \bibfield  {author} {\bibinfo {author} {\bibfnamefont {S.}~\bibnamefont
  {Ryu}}, \bibinfo {author} {\bibfnamefont {A.~P.}\ \bibnamefont {Schnyder}},
  \bibinfo {author} {\bibfnamefont {A.}~\bibnamefont {Furusaki}}, \ and\
  \bibinfo {author} {\bibfnamefont {A.~W.}\ \bibnamefont {Ludwig}},\ }\href
  {\doibase 10.1088/1367-2630/12/6/065010} {\bibfield  {journal} {\bibinfo
  {journal} {New Journal of Physics}\ }\textbf {\bibinfo {volume} {12}}
  (\bibinfo {year} {2010}),\ 10.1088/1367-2630/12/6/065010},\ \Eprint
  {http://arxiv.org/abs/0912.2157} {arXiv:0912.2157} \BibitemShut {NoStop}%
\bibitem [{\citenamefont {Akhmerov}\ \emph {et~al.}(2011)\citenamefont
  {Akhmerov}, \citenamefont {Dahlhaus}, \citenamefont {Hassler}, \citenamefont
  {Wimmer},\ and\ \citenamefont {Beenakker}}]{Akhmerov2011}%
  \BibitemOpen
  \bibfield  {author} {\bibinfo {author} {\bibfnamefont {A.~R.}\ \bibnamefont
  {Akhmerov}}, \bibinfo {author} {\bibfnamefont {J.~P.}\ \bibnamefont
  {Dahlhaus}}, \bibinfo {author} {\bibfnamefont {F.}~\bibnamefont {Hassler}},
  \bibinfo {author} {\bibfnamefont {M.}~\bibnamefont {Wimmer}}, \ and\ \bibinfo
  {author} {\bibfnamefont {C.~W.~J.}\ \bibnamefont {Beenakker}},\ }\href
  {\doibase 10.1103/PhysRevLett.106.057001} {\bibfield  {journal} {\bibinfo
  {journal} {Physical Review Letters}\ }\textbf {\bibinfo {volume} {106}},\
  \bibinfo {pages} {057001} (\bibinfo {year} {2011})},\ \Eprint
  {http://arxiv.org/abs/1009.5542} {arXiv:1009.5542} \BibitemShut {NoStop}%
\bibitem [{\citenamefont {Fulga}\ \emph {et~al.}(2011)\citenamefont {Fulga},
  \citenamefont {Hassler}, \citenamefont {Akhmerov},\ and\ \citenamefont
  {Beenakker}}]{Fulga2011}%
  \BibitemOpen
  \bibfield  {author} {\bibinfo {author} {\bibfnamefont {I.~C.}\ \bibnamefont
  {Fulga}}, \bibinfo {author} {\bibfnamefont {F.}~\bibnamefont {Hassler}},
  \bibinfo {author} {\bibfnamefont {A.~R.}\ \bibnamefont {Akhmerov}}, \ and\
  \bibinfo {author} {\bibfnamefont {C.~W.~J.}\ \bibnamefont {Beenakker}},\
  }\href {\doibase 10.1103/PhysRevB.83.155429} {\bibfield  {journal} {\bibinfo
  {journal} {Physical Review B}\ }\textbf {\bibinfo {volume} {83}},\ \bibinfo
  {pages} {155429} (\bibinfo {year} {2011})},\ \Eprint
  {http://arxiv.org/abs/1101.1749} {arXiv:1101.1749} \BibitemShut {NoStop}%
\bibitem [{\citenamefont {Groth}\ \emph {et~al.}(2014)\citenamefont {Groth},
  \citenamefont {Wimmer}, \citenamefont {Akhmerov},\ and\ \citenamefont
  {Waintal}}]{Kwant}%
  \BibitemOpen
  \bibfield  {author} {\bibinfo {author} {\bibfnamefont {C.~W.}\ \bibnamefont
  {Groth}}, \bibinfo {author} {\bibfnamefont {M.}~\bibnamefont {Wimmer}},
  \bibinfo {author} {\bibfnamefont {A.~R.}\ \bibnamefont {Akhmerov}}, \ and\
  \bibinfo {author} {\bibfnamefont {X.}~\bibnamefont {Waintal}},\ }\href
  {\doibase 10.1088/1367-2630/16/6/063065} {\bibfield  {journal} {\bibinfo
  {journal} {New Journal of Physics}\ }\textbf {\bibinfo {volume} {16}},\
  \bibinfo {pages} {063065} (\bibinfo {year} {2014})}\BibitemShut {NoStop}%
\bibitem [{\citenamefont {Novik}\ \emph {et~al.}(2005)\citenamefont {Novik},
  \citenamefont {Pfeuffer-Jeschke}, \citenamefont {Jungwirth}, \citenamefont
  {Latussek}, \citenamefont {Becker}, \citenamefont {Landwehr}, \citenamefont
  {Buhmann},\ and\ \citenamefont {Molenkamp}}]{Novik2005}%
  \BibitemOpen
  \bibfield  {author} {\bibinfo {author} {\bibfnamefont {E.~G.}\ \bibnamefont
  {Novik}}, \bibinfo {author} {\bibfnamefont {A.}~\bibnamefont
  {Pfeuffer-Jeschke}}, \bibinfo {author} {\bibfnamefont {T.}~\bibnamefont
  {Jungwirth}}, \bibinfo {author} {\bibfnamefont {V.}~\bibnamefont {Latussek}},
  \bibinfo {author} {\bibfnamefont {C.~R.}\ \bibnamefont {Becker}}, \bibinfo
  {author} {\bibfnamefont {G.}~\bibnamefont {Landwehr}}, \bibinfo {author}
  {\bibfnamefont {H.}~\bibnamefont {Buhmann}}, \ and\ \bibinfo {author}
  {\bibfnamefont {L.~W.}\ \bibnamefont {Molenkamp}},\ }\href {\doibase
  10.1103/PhysRevB.72.035321} {\bibfield  {journal} {\bibinfo  {journal} {Phys.
  Rev. B}\ }\textbf {\bibinfo {volume} {72}},\ \bibinfo {pages} {035321}
  (\bibinfo {year} {2005})}\BibitemShut {NoStop}%
\bibitem [{\citenamefont {Cardona}\ \emph {et~al.}(1986)\citenamefont
  {Cardona}, \citenamefont {Christensen},\ and\ \citenamefont
  {Fasol}}]{Cardona1986}%
  \BibitemOpen
  \bibfield  {author} {\bibinfo {author} {\bibfnamefont {M.}~\bibnamefont
  {Cardona}}, \bibinfo {author} {\bibfnamefont {N.~E.}\ \bibnamefont
  {Christensen}}, \ and\ \bibinfo {author} {\bibfnamefont {G.}~\bibnamefont
  {Fasol}},\ }\href {\doibase 10.1103/PhysRevLett.56.2831} {\bibfield
  {journal} {\bibinfo  {journal} {Phys. Rev. Lett.}\ }\textbf {\bibinfo
  {volume} {56}},\ \bibinfo {pages} {2831} (\bibinfo {year}
  {1986})}\BibitemShut {NoStop}%
\bibitem [{\citenamefont {Madelung}(2004)}]{Madelung2004}%
  \BibitemOpen
  \bibfield  {author} {\bibinfo {author} {\bibfnamefont {O.}~\bibnamefont
  {Madelung}},\ }\href@noop {} {\emph {\bibinfo {title} {{Semiconductors: Data
  Handbook}}}}\ (\bibinfo  {publisher} {Springer-Verlag Berlin Heidelberg},\
  \bibinfo {year} {2004})\BibitemShut {NoStop}%
\bibitem [{\citenamefont {Ashcroft}\ and\ \citenamefont
  {Mermin}(1976)}]{Ashcroft1976}%
  \BibitemOpen
  \bibfield  {author} {\bibinfo {author} {\bibfnamefont {N.~W.}\ \bibnamefont
  {Ashcroft}}\ and\ \bibinfo {author} {\bibfnamefont {N.~D.}\ \bibnamefont
  {Mermin}},\ }\href@noop {} {\emph {\bibinfo {title} {{S}olid {S}tate
  {P}hysics}}}\ (\bibinfo  {publisher} {Holt-Saunders},\ \bibinfo {year}
  {1976})\BibitemShut {NoStop}%
\bibitem [{\citenamefont {Fukui}\ \emph {et~al.}(2012)\citenamefont {Fukui},
  \citenamefont {Shiozaki}, \citenamefont {Fujiwara},\ and\ \citenamefont
  {Fujimoto}}]{Fukui2012}%
  \BibitemOpen
  \bibfield  {author} {\bibinfo {author} {\bibfnamefont {T.}~\bibnamefont
  {Fukui}}, \bibinfo {author} {\bibfnamefont {K.}~\bibnamefont {Shiozaki}},
  \bibinfo {author} {\bibfnamefont {T.}~\bibnamefont {Fujiwara}}, \ and\
  \bibinfo {author} {\bibfnamefont {S.}~\bibnamefont {Fujimoto}},\ }\href
  {\doibase 10.1143/JPSJ.81.114602} {\bibfield  {journal} {\bibinfo  {journal}
  {Journal of the Physical Society of Japan}\ }\textbf {\bibinfo {volume}
  {81}},\ \bibinfo {pages} {114602} (\bibinfo {year} {2012})},\ \Eprint
  {http://arxiv.org/abs/1206.4410} {arXiv:1206.4410} \BibitemShut {NoStop}%
\bibitem [{Note1()}]{Note1}%
  \BibitemOpen
  \bibinfo {note} {We estimate the coherence length in the SC regions to be
  $\xi =\protect \frac {\hbar v_F}{\Delta }=50.0\protect \tmspace +\thickmuskip
  {.2777em}\protect \mathrm {nm}$.}\BibitemShut {Stop}%
\bibitem [{\citenamefont {Hell}\ \emph {et~al.}(2017)\citenamefont {Hell},
  \citenamefont {Leijnse},\ and\ \citenamefont {Flensberg}}]{Hell2017}%
  \BibitemOpen
  \bibfield  {author} {\bibinfo {author} {\bibfnamefont {M.}~\bibnamefont
  {Hell}}, \bibinfo {author} {\bibfnamefont {M.}~\bibnamefont {Leijnse}}, \
  and\ \bibinfo {author} {\bibfnamefont {K.}~\bibnamefont {Flensberg}},\ }\href
  {\doibase 10.1103/PhysRevLett.118.107701} {\bibfield  {journal} {\bibinfo
  {journal} {Physical Review Letters}\ }\textbf {\bibinfo {volume} {118}},\
  \bibinfo {pages} {1} (\bibinfo {year} {2017})},\ \Eprint
  {http://arxiv.org/abs/1608.08769} {arXiv:1608.08769} \BibitemShut {NoStop}%
\end{thebibliography}%

\end{document}